# Universal Transport Dynamics of Complex Fluids

Sanggeun Song[1,2,3], Seong Jun Park[1,2,3], Bong June Sung[4], Jun Soo Kim[5], Ji-Hyun Kim*[1] and Jaeyoung Sung*[1,2,3]

[1] *Creative Research Initiative Center for Chemical Dynamics in Living Cells, Chung-Ang University, Seoul 06974, Republic of Korea*

[2] *Department of Chemistry, Chung-Ang University, Seoul 06974, Republic of Korea*

[3] *National Institute of Innovative Functional Imaging, Chung-Ang University, Seoul 06974, Republic of Korea*

[4] *Department of Chemistry, Sogang University, Seoul 04107, Republic of Korea*

[5] *Department of Chemistry and Nanoscience, Ewha Womans University, Seoul 03760, Republic of Korea*

**Author Information** The authors declare no competing financial interests. Correspondence and requests for materials should be addressed to J.S. (jaeyoung@cau.ac.kr) and J.-H.K. (jihyunkim@cau.ac.kr).

# ABSTRACT

Thermal motion in complex fluids is a complicated stochastic process but ubiquitously exhibits initial ballistic, intermediate sub-diffusive, and long-time non-Gaussian diffusive motion, unless interrupted. Despite its relevance to numerous dynamical processes of interest in modern science, a unified, quantitative understanding of thermal motion in complex fluids remains a long-standing problem. Here, we present a new transport equation and its solutions, which yield a unified quantitative explanation of the mean square displacement (MSD) and the non-Gaussian parameter (NGP) of various fluid systems. We find the environment-coupled diffusion kernel and its time correlation function are two essential quantities determining transport dynamics of complex fluids. From our analysis, we construct a general, explicit model of the complex fluid transport dynamics. This model quantitatively explains not only the MSD and NGP, but also the time-dependent relaxation of the displacement distribution for various systems. We introduce the concepts of intrinsic disorder and extrinsic disorder that have distinct effects on transport dynamics and different dependencies on temperature and density. This work presents a new paradigm for quantitative understanding of transport and transport-coupled processes in complex disordered media.

Thermal motion in complex fluids is a complex stochastic process, which underlies a diverse range of dynamical processes of interest in modern science. Although thermal motion in condensed media has been the subject of a great deal of research since Einstein's seminal work on Brownian motion[1] and is the most mature topic in non-equilibrium statistical physics, it still remains a daunting task to achieve a quantitative understanding of the transport dynamics and the displacement distribution in disordered fluidic systems, including cell nuclei and cytosols[2], membranes and biological tissue[3], polymeric fluid[4], supercooled water [5, 6], ionic liquids[7, 8], and dense hard-disc fluids[9]. Interestingly, thermal motion of diverse complex fluid systems commonly exhibits time-dependent phase changes in the mean square displacement, from initial ballistic dynamics, to intermediate sub-diffusive dynamics, and then to terminal diffusion[10, 11, 12]; the associated displacement distribution is non-Gaussian except in the short and long time limits, and its deviation from Gaussian increases at short times but decreases at long times, as long as thermal motion is uninterrupted. These phenomena cannot be quantitatively explained by the original theory of Brownian motion[13].

The theory of Brownian motion has undergone a number of important generalizations[14, 15, 16, 17, 18, 19, 20, 21, 22, 23, 24]. Among these generalizations, the continuous time random walk (CTRW) model proposed by Montroll and Weiss[20] enables quantitative description of anomalous transport caused by tracer particle being trapped by other objects in media and non-classical kinetics of chemical reactions in disordered media[25, 26, 27]; the CTRW with a power-law waiting time distribution (WTD), $\psi(t) \propto t^{-(1+\alpha)}$ $(0 < \alpha < 1)$, successfully explains charge transport in amorphous semiconductors[28], which shows a sub-diffusive power-law mean square displacement. This sub-diffusive transport dynamics can be described by the fractional diffusion equation or the fractional Fokker-Planck equation[29, 30, 31] in the continuum limit. Fractional Brownian motion (FBM) proposed by Mandelbrot and Ness[21] models anomalous diffusion due to anti-persistent motion in a viscoelastic medium; FBM is a Gaussian sub-diffusive process, while the CTRW with a power-law WTD is a non-Gaussian process. O'Shaughnessy and Procaccia, and Havlin and Ben-Avraham investigate anomalous transport originating from self-similarity of transport media, considering a random walk or diffusion in a fractal[23, 24]. Recently, Novikov *et al.*, introduced a model of anomalous thermal motion dependent on mesoscopic structures of media and analyze the long-time behavior of the diffusion coefficient by using the renormalization group solution[32, 33]. Despite their applicability to a number of natural phenomena[28, 32, 34, 35], the predictions of these models are not consistent with Fickian yet non-Gaussian diffusion ubiquitously observed in disordered fluidic systems[36, 37, 38, 39, 40, 41]. This issue was recently addressed by stochastic diffusivity (SD) models, in which the diffusion coefficient is treated as a stochastic variable[42, 43, 44, 45]. While SD models successfully demonstrate Fickian yet non-Gaussian diffusion, to the best of our knowledge, these models are inconsistent with the transient sub-diffusive mean square displacement (MSD) and non-monotonic time-dependence of the non-Gaussian parameter (NGP), widely observed features of complex fluids.

**Transport equation of complex fluids**

Here, to solve this challenging problem, we present a new transport equation that provides an accurate quantitative description of thermal motion for various complex fluid systems. This equation can be derived by considering the continuum limit of a random walk model with a general sojourn time distribution, $\psi_\Gamma(t)$, that is coupled to arbitrary hidden environmental variables $\boldsymbol{\Gamma}$ (See Fig.1). In this model, hidden environmental variables designate the entire set of dynamical variables that affect transport dynamics in disordered fluids. Our transport equation reads as

$$\hat{\dot{p}}(\mathbf{r},\Gamma,s) = \hat{\mathcal{D}}_\Gamma(s)\nabla^2 \hat{p}(\mathbf{r},\Gamma,s) + L(\Gamma)\hat{p}(\mathbf{r},\Gamma,s) \tag{1}$$

where $\hat{\dot{p}}(\mathbf{r},\Gamma,s)$ and $\hat{p}(\mathbf{r},\Gamma,s)$ denote the Laplace transform of $\partial p(\mathbf{r},\Gamma,t)/\partial t$ and $p(\mathbf{r},\Gamma,t)$, respectively; $p(\mathbf{r},\Gamma,t)$ denotes the joint probability density that a particle is at position $\mathbf{r}$ and the hidden environment is at state $\Gamma$ at time $t$. This joint probability density satisfies the following normalization condition: $\int d\mathbf{r} \int d\Gamma\, p(\mathbf{r},\Gamma,t) = 1$. Throughout this work, $\hat{f}(s)$ denotes $\int_0^\infty dt\, e^{-st} f(t)$. In equation (1), $\hat{\mathcal{D}}_\Gamma(s)$ designates the diffusion kernel defined by $\hat{\mathcal{D}}_\Gamma(s) = \varepsilon^2 \hat{\kappa}_\Gamma(s)/2d$ with $\hat{\kappa}_\Gamma(s) \equiv s\hat{\psi}_\Gamma(s)/[1-\hat{\psi}_\Gamma(s)]$, where $\varepsilon$ and $d$ respectively denote the lattice constant and the spatial dimension. $\hat{\kappa}_\Gamma(s)$ is dependent on $\varepsilon$; for the continuum limit description, we assume $\lim_{\varepsilon \to 0} \varepsilon^2 \hat{\kappa}_\Gamma(s)$ exists. $L(\Gamma)$ designates a mathematical operator describing the dynamics of the hidden environmental variables $\Gamma$. For the sake of generality, we do not assume a particular model of environmental state dynamics, nor do we assume a particular form of mathematical operator $L(\Gamma)$. A correct mathematical form of $L(\Gamma)$ is dependent on the environment surrounding the system in question; when environmental state dynamics is a non-Markov process, $L(\Gamma)$ may be dependent on Laplace variable $s$. As demonstrated in this work, robust and quantitative information about transport dynamics coupled to hidden environmental variables can be extracted from simultaneous analysis of the MSD and NGP time profiles. This information can then be used to construct a more explicit model of transport dynamics of complex fluidic systems. Equation (1) encompasses the CTRW model, the diffusing diffusivity models, and their various generalizations.

We present the derivation of equation (1) in Methods. A further generalization of equation (1) for complex fluidic systems under an external potential is presented in Supplementary Note 1 of Supplementary Information.

**Analytic Expressions of the Moments**

From equation (1), we obtain the exact analytical expressions of the first two non-vanishing moments, $\langle |\mathbf{r}(t) - \mathbf{r}(0)|^2 \rangle \, (\equiv \Delta_2(t))$ and $\langle |\mathbf{r}(t) - \mathbf{r}(0)|^4 \rangle \, (\equiv \Delta_4(t))$, of the displacement distribution:

$$\hat{\Delta}_2(s) = \frac{2d}{s^2} \langle \hat{\mathcal{D}}_\Gamma(s) \rangle \tag{2a}$$

$$\hat{\Delta}_4(s) = \left(1 + \frac{2}{d}\right) 2s \hat{\Delta}_2(s)^2 \left[1 + s\hat{C}_\mathcal{D}(s)\right] \tag{2b}$$

In obtaining equations (2a) and (2b), we assume that the hidden environment reaches a stationary state, such as the equilibrium state or the nonequilibrium steady-state, at long times. The bracket notation here designates the average over the stationary distribution of environmental states.

The average diffusion kernel, $\langle \mathcal{D}_\Gamma(t) \rangle$, in equation (2a) is nothing but the velocity autocorrelation function, i.e., $\langle \mathcal{D}_\Gamma(t) \rangle = \langle \mathbf{v}(t) \cdot \mathbf{v}(0) \rangle / d$ with $\mathbf{v}(t)$ being the velocity vector. This can be clearly seen by comparing equation (2a) and the Laplace transform of the well-known relation, $\Delta_2(t) = 2 \int_0^t d\tau_2 \int_0^{\tau_2} d\tau_1 \langle \mathbf{v}(\tau_2 - \tau_1) \cdot \mathbf{v}(0) \rangle$ [46], valid for any stationary transport process. Knowing this and utilizing the Tauberian theorem, we obtain $\lim_{s \to \infty} s \langle \hat{\mathcal{D}}_\Gamma(s) \rangle = \lim_{t \to 0} \langle \mathbf{v}(t) \cdot \mathbf{v}(0) \rangle / d = \langle |\mathbf{v}|^2 \rangle / d = k_B T / M$ with $k_B T$ and $M$ denoting thermal energy and the mass of the particle, respectively. This means that $\langle \hat{\mathcal{D}}_\Gamma(s) \rangle$ is proportional to the mean square velocity in the large $s$ limit, i.e., $\langle \hat{\mathcal{D}}_\Gamma(s) \rangle \cong s^{-1} \langle |\mathbf{v}|^2 \rangle / d$ $(s \to \infty)$. On the other hand, in the small $s$ limit, the value of $\langle \hat{\mathcal{D}}_\Gamma(s) \rangle$ approaches $\langle \hat{\mathcal{D}}_\Gamma(0) \rangle = \int_0^\infty dt \langle \mathbf{v}(t) \cdot \mathbf{v}(0) \rangle / d$, which is simply the diffusion constant, $\bar{D}$, according to the Green-Kubo relation [47]. Substituting the small (large) $s$ limit asymptotic behavior of $\langle \hat{\mathcal{D}}_\Gamma(s) \rangle$ into equation (2a), we recover the well-known asymptotic behavior of the MSD: $d(k_B T/M)t^2$ at short times and $2d\bar{D}t$ at long times.

While the second moment, $\Delta_2(t)$, is dependent only on the averaged diffusion kernel, $\langle \mathcal{D}_\Gamma(t) \rangle$, the fourth moment, $\Delta_4(t)$, is dependent on the environment-coupled fluctuation of the diffusion kernel, $\mathcal{D}_\Gamma(t)$. In equation (2b), $\hat{C}_\mathcal{D}(s)$ is the Laplace transform of the time correlation function (TCF) of the diffusion kernel fluctuation (see Methods for the precise definition). At long times where the MSD is linear in time, the diffusion kernel becomes the diffusion coefficient, i.e., $\hat{\mathcal{D}}_\Gamma(s) \cong \hat{\mathcal{D}}_\Gamma(0) (\equiv D_\Gamma)$ so that $C_\mathcal{D}(t)$ can be identified as the TCF of the diffusion coefficient fluctuation, i.e., $C_\mathcal{D}(t) \cong \langle \delta D(t) \delta D(0) \rangle / \langle D \rangle^2 = \eta_D^2 \phi_D(t)$. Throughout this work, $\eta_q^2$ and $\phi_q(t)$ ($q \in \{v^2, D\}$) designate the relative variance, $\langle \delta q^2 \rangle / \langle q \rangle^2$, and the normalized time correlation function of $q$, defined by $\phi_q(t) = \langle \delta q(t) \delta q(0) \rangle / \langle \delta q^2 \rangle$. At short times, on the other hand, $C_\mathcal{D}(t)$ can be identified as the TCF of squared speed $v^2(t) (\equiv |\mathbf{v}(t)|^2)$, i.e., $C_\mathcal{D}(t) \cong d\eta_{v^2}^2 \phi_{v^2}(t)$ $\left[ = d \langle \delta v^2(t) \delta v^2(0) \rangle / \langle v^2 \rangle^2 \right]$ (see Methods). Given that the initial speed distribution obeys the Maxwell-Boltzmann distribution, we obtain $\langle v^4 \rangle = (1 + 2/d) \langle v^2 \rangle^2$, and the initial value of $C_\mathcal{D}(t)$ can then only be given by $\lim_{t \to 0} C_\mathcal{D}(t) = d(\langle v^4 \rangle - \langle v^2 \rangle^2) / \langle v^2 \rangle^2 = 2$. We find this is true for both the supercooled water and the dense hard disc fluids investigated in the current work (see Fig. 2d and Supplementary Fig. 1). In our theory, $C_\mathcal{D}(t)$ is the essential function that characterizes environment-coupled fluctuation of the transport dynamics of complex fluids. When $C_\mathcal{D}(t) = 0$, equations (2a) and (2b) reduce to the results of the CTRW model in the continuum limit.

Using equations (2a) and (2b) and the definition of NGP, $\alpha_2(t)\left[=\Delta_4(t)/\left[(1+2/d)\Delta_2(t)^2\right]-1\right]$, we can quantitatively explain the MSD and NGP of thermal motion of various complex fluids. In Fig. 2, we demonstrate our quantitative analysis of molecular dynamics (MD) simulation results of the MSD and NGP for supercooled water. From the quantitative analysis, we extract the time profiles of $\langle \mathcal{D}_\Gamma(t) \rangle$ and $C_\mathcal{D}(t)$. This quantitative information is robust and model-independent, because the time profiles are extracted without assuming a particular model for the hidden environment or its influence over the diffusion kernel. As shown later in this work, this information is useful in constructing an explicit model of transport dynamics of complex fluid systems; from this explicit model, we can predict or quantitatively understand the time-dependence of the displacement distribution.

It is worth mentioning that the long-time limit value of $C_\mathcal{D}(t)$ is equal to the long-time limiting value of the NGP, i.e., $C_\mathcal{D}(\infty) = \alpha_2(\infty)$, which can be shown by using equations (2a) and (2b) and the definition of the NGP (see Methods). This result shows that $\alpha_2(\infty)$ remains finite only for a non-ergodic system, for which the TCF of the diffusion coefficient does not vanish even in the long-time limit. Therefore, the long-time limit value of the NGP can serve as an ergodicity measure for transport systems (see Supplementary Fig. 2).

**Intrinsic and Extrinsic disorder**

For an ergodic system, the long-time limit value of the product between the MSD and NGP can serve as a measure of disorder strength for complex fluids and is decomposable into intrinsic disorder, originating from non-Poisson *mean* transport dynamics, and extrinsic disorder, originating from environment-coupled *fluctuation* in transport dynamics. To show this, let us examine the long-time asymptotic behavior of the MSD and NGP:

$$\Delta_2(t) \cong 2d\bar{D}t + \Delta_c \qquad (t \to \infty) \tag{3a}$$

$$\alpha_2(t) \cong \frac{2\hat{C}_\mathcal{D}(0) + \Delta_c/d\bar{D}}{t} \qquad (t \to \infty) \tag{3b}$$

Equation (3a) can be obtained by substituting the Maclaurin series of $\langle \hat{\mathcal{D}}_\Gamma(s) \rangle$, $\langle \hat{\mathcal{D}}_\Gamma(s) \rangle = (\varepsilon^2/2d)\langle \hat{\kappa}_\Gamma(s) \rangle = (\varepsilon^2/2d)[\langle \hat{\kappa}_\Gamma(s) \rangle + \langle \hat{\kappa}'_\Gamma(s) \rangle s + \cdots]$, into equation (2a) and by taking the inverse Laplace transform of the resulting equation. In equation (3a), $\Delta_c$ emerges whenever the transport dynamics deviates from a simple Poisson process, or whenever the environment-coupled sojourn time distribution, $\psi_\Gamma(t)$, deviates from an exponential function (see Supplementary Note 2 in the Supplementary Information). We present the derivation of equations (3a) and (3b) in Methods. From these equations, we obtain the long-time limit value of the product of the MSD and NGP as

$$\lim_{t \to \infty} \langle r^2(t) \rangle \alpha_2(t) = 2\left[\Delta_c + 2d\bar{D}\hat{C}_\mathcal{D}(0)\right] \tag{4}$$

We define the disorder strength of complex fluids as $\lim_{t\to\infty} \langle r^2(t)\rangle \alpha_2(t)/\sigma^2$, with $\sigma$ being the effective diameter of a tracer particle. Equation (4) tells us that disorder strength has two different sources: $2\Delta_c/\sigma^2$ and $4d\bar{D}\hat{C}_D(0)/\sigma^2$, which originate from the non-Poisson *mean* transport dynamics and from the environment-coupled *fluctuation* in transport dynamics, respectively. We designate the latter term extrinsic disorder, which is quite sensitive to temperature and density of the environment, as demonstrated in Fig. 2e and Supplementary Fig. 1e. On the other hand, we designate $2\Delta_c/\sigma^2$ intrinsic disorder because this term persists even when environment-coupled fluctuation in transport dynamics is negligible. Intrinsic disorder is far less sensitive to the temperature and density of media than extrinsic disorder and can be easily estimated from equation (3a), the asymptotic long-time behavior of the MSD. Extrinsic disorder can be estimated by a direct numerical calculation of $4d\bar{D}\hat{C}_D(0)/\sigma^2$ or, more simply, by subtracting intrinsic disorder from total disorder strength (Fig. 2e).

**Quantitative Analysis of MSD and NGP**

Intrinsic disorder, or non-Poisson *mean* transport dynamics, causes the MSD to deviate from $2d\bar{D}t$, the prediction of the simple diffusion equation. We find that the following formula provides a quantitative description of the entire range of the MSD time profiles for various disordered fluids (see Supplementary Note 3 in the Supplementary Information):

$$\Delta_2(t) = 2d\frac{k_BT}{M\gamma_0^2}c_0(\gamma_0 t - 1 + e^{-\gamma_0 t}) + 2d\frac{k_BT}{M}\sum_{i=1}^{n}\frac{c_i}{\omega_{0,i}^2}\left[1 - e^{-\gamma_i t}\left(\cosh\omega_i t + \frac{\gamma_i}{\omega_i}\sinh\omega_i t\right)\right] \quad (5)$$

This equation represents the MSD of a bead in a polymer, but quantitatively explains the MSD of liquid water and dense hard disc fluids as well, as shown in Fig. 2a and Supplementary Fig. 1a. The applicability of equation (5) to various disordered fluid systems implies a universality in the transport dynamics of disordered fluids, which is decomposable into an unbound mode dynamics and multiple bound mode dynamics, comparable to viscoelastic motion of a bead in a polymer network. At short times, a tracer molecule is in a bound state, trapped by the surrounding molecules. This bound state consists of multiple bound modes with their own characteristic frequencies. At long times, on the other hand, a tracer molecule escapes the cage of the surrounding molecules and moves around in the media, repeatedly being caged and escaping the cage. The first term on the right-hand side of equation (5) accounts for the contribution from the unbound mode, and the second term accounts for the contribution from the bound modes. In equation (5), $c_i$ and $\gamma_i$ designate the weight coefficient and relaxation rate of the $i$th mode ($0 \leq i \leq n$). The weight coefficients are normalized by $\sum_{i=0}^{n} c_i = 1$. $\omega_{0,i}$ is the natural frequency of the $i$th bound mode and is related to $\omega_i$ as $\omega_i = \sqrt{\gamma_i^2 - \omega_{0,i}^2}$. At all times, temperatures, and densities investigated, equation (5) with only two bound modes ($n = 2$) already provides an accurate, quantitative explanation of the simulation results for the anomalous MSD of supercooled water and hard disc fluids (see Figs. 2a and Supplementary Fig. 1a), and the experimental results for colloidal particles moving along lipid tubes (see Fig. 4)[36]. According to equation (2a), the analytic expression of diffusion kernel yielding the MSD given in equation (5) can be obtained by $\langle \mathcal{D}_\Gamma(t) \rangle = \partial^2 \Delta_2(t)/\partial t^2$.

The NGP is dependent not only on the *mean* transport dynamics, or $\langle \hat{\mathcal{D}}_\Gamma(s) \rangle$, but also on *fluctuation* in transport dynamics, or $C_\mathcal{D}(t)$. For a simple diffusion process, we have $\langle \hat{\mathcal{D}}_\Gamma(s) \rangle = D$ and $C_\mathcal{D}(t) = 0$, and equations (2a) and (2b) yield $\Delta_2(t) = 2dDt$ and $\Delta_4(t) = (1+2/d)\Delta_2(t)^2$ so that the NGP vanishes. However, whenever $\langle \mathcal{D}_\Gamma(t) \rangle$ is not constant and/or $C_\mathcal{D}(t) \neq 0$, the NGP does not vanish. We find that, for disordered fluid systems investigated in this work, the time profiles of the NGP cannot be quantitatively understood when we assume $C_\mathcal{D}(t) = 0$, or fluctuations in the transport dynamics are absent (Supplementary Fig 3).

The NGP of disordered fluids is a non-monotonic function of time with a single peak. We find that the NGP quadratically increases with time, $\alpha_2(t) \propto t^2$, at short times (Supplementary Note 4) but decreases with time, $\alpha_2(t) \propto t^{-1}$, at long times following equation (3b). As shown in Fig. 2a it is only after the NGP peak time that Fickian diffusion emerges. These properties of the NGP are not specific to supercooled water but common across various disordered fluids [9].

The NGP peak height, $\alpha_2(\tau_{ng})$, serves as a measure of the relative variance of the diffusion coefficient for disordered fluids. From the displacement distribution at the NGP peak time, we can extract the distribution of the diffusion coefficient using the method proposed in ref. [37]. We find the relative variance, $\eta_D^2$, of the extracted diffusion coefficient distribution has the same value as the NGP peak height (see Supplementary Fig. 4 in the Supplementary Information). This is not a coincidence. We can show that the NGP peak height has the same value as the relative variance of the diffusion coefficient at the Fickian diffusion onset time, or the NGP peak time, $\tau_{ng}$ (Methods). Both the NGP peak height and the NGP peak time increase with inverse temperature and density, as demonstrated in Fig. 2b and Supplementary Fig. 1b.

**Construction of an Explicit Model**

In order to provide a quantitative explanation of the time profile of $C_\mathcal{D}(t)$ extracted from the MSD and NGP, we construct an explicit model of the environment-coupled diffusion kernel. By comparing equation (2a) and Laplace transform of the MSD given in equation (5), we obtain the following expression: $\langle \hat{\mathcal{D}}_\Gamma(s) \rangle = c_0 \hat{f}_0(s) + \sum_{i=1}^n c_i \hat{f}_i(s)$, where $\hat{f}_0(s)$ and $\hat{f}_i(s)$ denote the diffusion kernels associated with the unbound mode dynamics and the $i$th bound mode dynamics, given by $\hat{f}_0(s) = (k_B T/M)(s+\gamma_0)^{-1}$ and $\hat{f}_i(s) = (k_B T/M)s\left[(s+\gamma_i)^2 - \omega_i^2\right]^{-1}$, respectively. We can extend this equation by assuming the weight coefficients $\{c_0, c_1, \cdots, c_n\}$ are dependent on environmental state variables $\Gamma$, obtaining the following model of the diffusion kernel:

$$\hat{\mathcal{D}}_\Gamma(s) = c_0(\Gamma)\hat{f}_0(s) + \sum_{i=1}^n c_i(\Gamma)\hat{f}_i(s) \tag{6}$$

From this model, we obtain the following analytic expression for $\hat{C}_\mathcal{D}(s)$ (see Methods):

$$\hat{C}_\mathcal{D}(s)\langle\hat{\mathcal{D}}_\Gamma(s)\rangle^2 = \frac{\gamma_0^2 \langle \delta D^2\rangle \hat{\phi}_D(s)}{(s+\gamma_0)^2} + \sum_{i=0}^n \sum_{j=0}^n{}' \frac{\zeta_{ij}\hat{f}_i(s)\hat{f}_j(s)}{s+\lambda_{ij}} \tag{7}$$

where the prime notation in the second term signifies that the sum excludes the term with $i = j = 0$. In the derivation of equation (7), we assume the time correlation between weight coefficients is given by $\langle \delta c_i(t) \delta c_j(0) \rangle = \zeta_{ij} \exp(-\lambda_{ij} t)$ with $\zeta_{ij}$ and $\lambda_{ij}$ being constant in time, unless $i$ and $j$ are zero. Noting that, in equation (6), $\hat{\mathcal{D}}_\Gamma(s) \cong c_0(\Gamma)(k_B T/M\gamma_0)$ in the small $s$ regime, $s \ll \gamma_i$, we can relate the weight coefficient TCF, $\langle \delta c_0(t) \delta c_0(0) \rangle$, of the unbound mode to the TCF of the diffusion coefficient fluctuation by $\langle \delta c_0(t) \delta c_0(0) \rangle = \langle \delta D(t) \delta D(0) \rangle (M\gamma_0/k_B T)^2$ at times longer than any element of $\{\gamma_i^{-1}\}$, whose magnitudes are less than 1 picosecond for supercooled water (see Table 1). In the small $s$ regime, only the first term on the R.H.S. equation (7) contributes to the relaxation of diffusion kernel fluctuation, which leaves us with $C_\mathcal{D}(t) \cong \phi_\mathcal{D}(t)\eta_\mathcal{D}^2$. This result shows that diffusion kernel correlation becomes the TCF of the diffusion coefficient at long times. Recalling that $\eta_\mathcal{D}^2 \cong \alpha_2(\tau_{ng})$, we then extract $\phi_\mathcal{D}(t)$ from the time-profile of $C_\mathcal{D}(t)/\alpha_2(\tau_{ng})$ at times longer than the NGP peak time. The long-time tail of $C_\mathcal{D}(t)$ or $\phi_\mathcal{D}(t)$ extracted from the MSD and NGP can be explained by an explicit model of the diffusion coefficient fluctuation described in the next section. With this model, equation (7) provides a quantitative explanation of the $C_\mathcal{D}(t)$ time profile of supercooled water for the entire time range (see Fig. 2f).

**Quantitative Explanation of Fickian yet non-Gaussian Diffusion**

Disordered fluids exhibit Fickian yet non-Gaussian diffusion, where the displacement distribution is strongly non-Gaussian even at long times where the MSD linearly increases with time[9, 10, 36, 37, 38]. Put briefly, the displacement distribution of disordered fluids starts as Gaussian with variance given by $t^2 k_B T/M$ at short times but deviates from Gaussian at any finite time. At the NGP peak time, where the deviation from Gaussian is greatest, the displacement distribution appears similar to a Laplace distribution with an exponential tail. It is at this time that the non-Gaussian displacement distribution begins its relaxation to Gaussian and the MSD becomes linear in time. This phenomenon is widely observed across various disordered fluid systems[7, 9, 10], but has yet to be quantitatively explained.

To understand the time-dependent relaxation of the non-Gaussian displacement distribution in the Fickian diffusion regime, we need an explicit model of the diffusion coefficient fluctuation for the fluid system in question. In the literature, the diffusion coefficient is often modelled as $D = A\exp(-\beta E)$, where $A$, $\beta$, and $E$ are the entropic factor, inverse thermal energy, and activation energy, respectively[48]. This model can be generalized by assuming that the entropic factor and activation energy can be stochastic variables dependent on hidden environmental variables. In this work, we make this generalization and consider two exactly solvable models. The first model assumes that the diffusion coefficient is given by $D_\Gamma = A\exp(-\beta E_\Gamma)$ where the fluctuation of $E_\Gamma$ around its mean value, $\langle E_\Gamma \rangle$, is given by $E_\Gamma(t) - \langle E \rangle = \sum_k b_k \Gamma_k(t)$, where $\{b_k\}$ and $\{\Gamma_k(t)\}$ are constants and stationary Gaussian Markov processes, also known as Ornstein-Uhlenbeck (OU) processes (see Methods)[49]. For this model, an exact analytic expression of $\phi_D(t)$ is available (see Methods and Figure 3). Combined with the analytic expression of $\phi_D(t)$, equation (7) provides a quantitative explanation of $C_D(t)$ for supercooled water, as shown in Figure 2f. The optimized parameter values are presented in Table 1. Using the first model with optimized parameter values, we can predict the time-dependent relaxation of the non-Gaussian displacement distribution in the Fickian diffusion regime (see Methods). The prediction of this model is in excellent agreement with the MD simulation results for supercooled water, as shown in Fig. 3b. In the second model, we model the diffusion coefficient as $D_\Gamma(t) = A_\Gamma(t)\exp(-\beta E_\Gamma(t)) \cong \langle D \rangle \sum_k a_k \Gamma_k^2(t)$ where $\{a_k\}$ and $\{\Gamma_k(t)\}$ are constants and OU processes, respectively. This model is a generalization of the model in ref. [44] and yields an analytic expression of the displacement distribution (see equation (53) in Methods). We find this expression provides an excellent quantitative explanation of the experimentally measured displacement distribution of colloidal beads diffusing on lipid tubes reported in ref. [36] (See Fig. 3d). The optimized parameters of the second model are also presented in Table 1. Equation (7) with optimized parameter values allows us to calculate the time profiles of the MSD and the diffusion kernel correlation for the colloidal bead system (see Supplementary Fig. 5).

The displacement distribution approaches Gaussian only after individual displacement trajectories become statistically equivalent. If individual displacement trajectories are statistically equivalent, the ergodicity breaking (EB) parameter proposed by He, Burov, Metzler, and Barkai[50] is linear in $t/t_{max}$, where $t$ and $t_{max}$ denote the time lag, or the interval over which the time-averaged MSD is calculated, and the maximum trajectory length, respectively[51]. Otherwise, the EB parameter deviates from its linear dependence on $t/t_{max}$. As shown in Fig. 4a, at temperatures lower than 230 K, the EB parameter of supercooled water shows anomalous power-law dependence on $t/t_{max}$ at short times but resumes normal linear dependence on $t/t_{max}$ at times longer than characteristic time $\tau_{EB}$. Deviation of the displacement distribution from Gaussian, measured by the NGP, becomes negligible only at times much longer than $\tau_{EB}$, as demonstrated in Fig. 4b for supercooled water. On the other hand, $C_\mathcal{D}(t)$ is negligibly small at characteristic time $\tau_{EB}$ (see Fig. 4b). This is due to the fact that the long-time relaxation of the NGP is not only contributed from extrinsic disorder leading to the trajectory-to-trajectory variation in the transport dynamics, but also from intrinsic disorder, whose effects persist even for homogeneous systems with statistically equivalent displacement trajectories (see equation (3b)). This analysis shows that the long-time tail of $C_\mathcal{D}(t)$ better characterizes the relaxation of diffusivity fluctuation than the NGP[52].

**Outlook**

The essential feature of our approach to transport dynamics of complex fluids is hidden environmental variables that represent the entire set of dynamic variables affecting transport dynamics of our tracer particles. By accounting for their effects without using an *a priori* explicit model, this approach enables the extraction of robust, quantitative information about the transport dynamics of complex fluid systems. This information can then be used to construct a more explicit model of transport dynamics of the environment-coupled system in question. In achieving quantitative understanding of complex systems, this type of approach is advantageous over the conventional approach that relies on fully explicit models of the system, the environment, and their interactions. This is because, for a system interacting with a complex environment, it is difficult to construct a model that is both accurate and explicit from the onset, due to lack of information. Our approach is applicable to quantitative investigation of various other complex systems in natural science[53, 54]. We leave further applications of this work for future research.

**Acknowledgements**

The authors gratefully acknowledge Professors Steve Granick, Eli Barkai, Ralf Metzler, and Jae-Hyung Jeon for their helpful comments, and Mr. Luke Bates for his careful reading of our manuscript. This work was supported by the Creative Research Initiative Project program (2015R1A3A2066497) and the NRF grant (MSIP) (2015R1A2A1A15055664) funded by the Korean government. S. Song would also like to acknowledge the Chung-Ang University Graduate Research Scholarship in 2016.

**Author contributions**

S.S., S.J.P., J.-H.K., and J.S. provided the new model and method of analysis; S.S. and J.-H.K. performed analysis; S.S., B.J.S., and J.S.K. performed computer simulations; S.S., B.J.S., J.S.K., J.-H.K., and J.S. wrote the manuscript; J.-H.K. and J.S. designed the research. J.S. supervised the entire research project.

# Figures

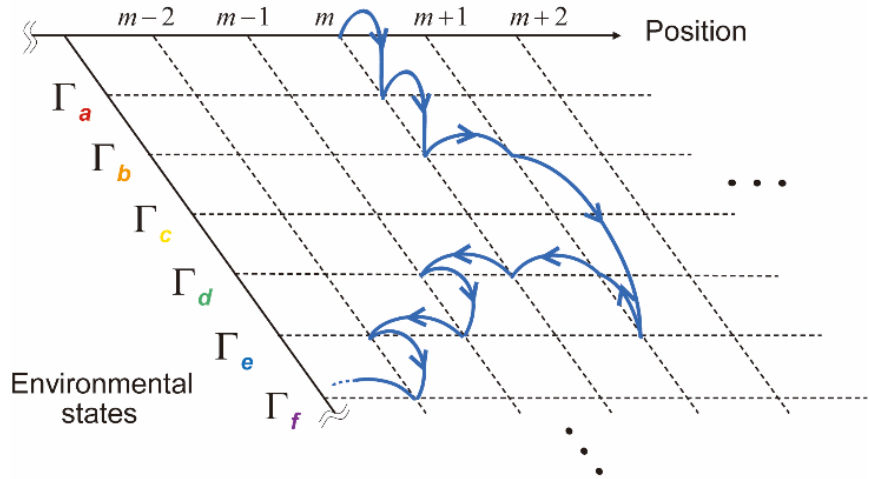

**FIG. 1 | Schematic representation of the current random walk model with environmental state dependent dynamics.** In our model, the sojourn time distribution $\psi_\Gamma(t)$ of a random walker at each point is dependent on environmental state variables. $\Gamma$ represents an entire set of dynamic variables that affects transport dynamic or sojourn time distribution of the random walker. Any assumption about dynamics of $\Gamma$ and its coupling to the sojourn time distribution is not assumed. The probabilistic dynamics of this random walker model can be described by equation (1) in the continuum limit.

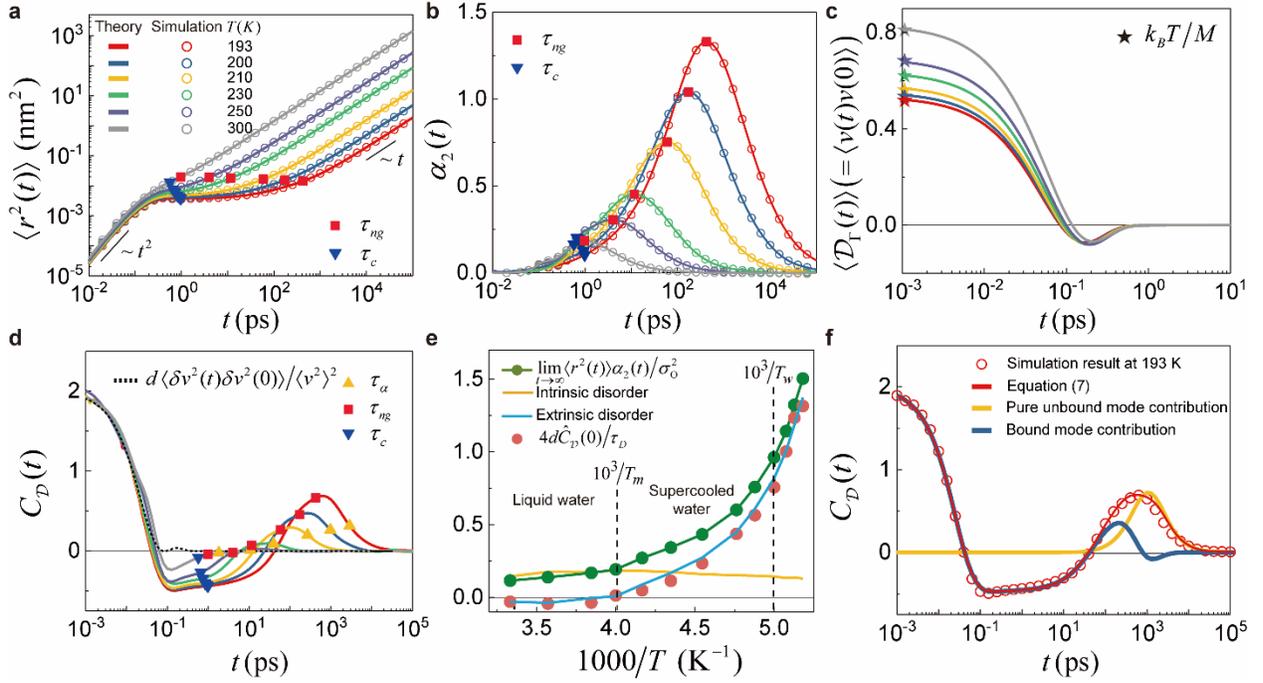

**FIG. 2 | Quantitative analysis of the MSD and NGP for the TIP4P/2005 water system. a, b**, MSDs and NGPs as functions of time at various temperatures. The solid lines represent the best least-square fits of equation (5) and a linear combination of three or four Gaussian-shaped functions to the simulation results (open circles), respectively. **c,** Averaged diffusion kernel, or equivalently the velocity autocorrelation function, obtained from the second-order time derivatives of the best MSD fits given in **a**. **d**, The diffusion kernel correlation function, $C_\mathcal{D}(t)$, obtained by analyzing the MSD and NGP (see Methods). The dotted line represents the mean-scaled time correlation function of squared speed fluctuation at 193 K, which is the short-time approximation of $C_\mathcal{D}(t)$. In **a**, **b**, and **d**, the navy-blue triangles and the red squares represent the caging times, $\tau_c$, and the NGP peak times, $\tau_{ng}$, respectively. In **d**, the yellow triangles represent the alpha relaxation times, $\tau_\alpha$, estimated from the intermediate scattering function (see Supplementary Fig. 6). **e,** Total disorder, $\lim_{t\to\infty}\langle r^2(t)\rangle\alpha_2(t)/\sigma_O^2$, scaled by an oxygen atom's Lennard-Jones diameter

squared, $\sigma_O^2$ ($= 3.1589^2$ Å$^2$) (green circles) and its two components: intrinsic and extrinsic disorder (yellow and cyan lines) (see text below equation (4)). Extrinsic disorder, estimated from the difference between the total disorder and intrinsic disorder, is in good agreement the value of $4d\hat{C}_\mathcal{D}(0)/\tau_D$ (red circles), where $d$, $\hat{C}_\mathcal{D}(0)$, and $\tau_D \left(= \sigma_O^2/\overline{D}\right)$ respectively denote the spatial dimension, the whole-time integration of $C_\mathcal{D}(t)$ given in **d**, and the diffusion time scale. $T_m$ and $T_W$ denote the melting temperature[55] and the Widom line temperature[56] at which a crossover between fragile supercooled water and strong supercooled water occurs[57] (see Supplementary Fig. 7). **f.** The time profile of $C_\mathcal{D}(t)$ for supercooled water at 193 K. (red circles) Results extracted from the simulation results of MSD and NGP, (red line) best-fitted result of equation (7), (yellow line) the contribution of the pure unbound mode, or the first term on the R.H.S. of equation (7), (blue line) the contribution of the bound modes, or the second term on the R.H.S. of equation (7).

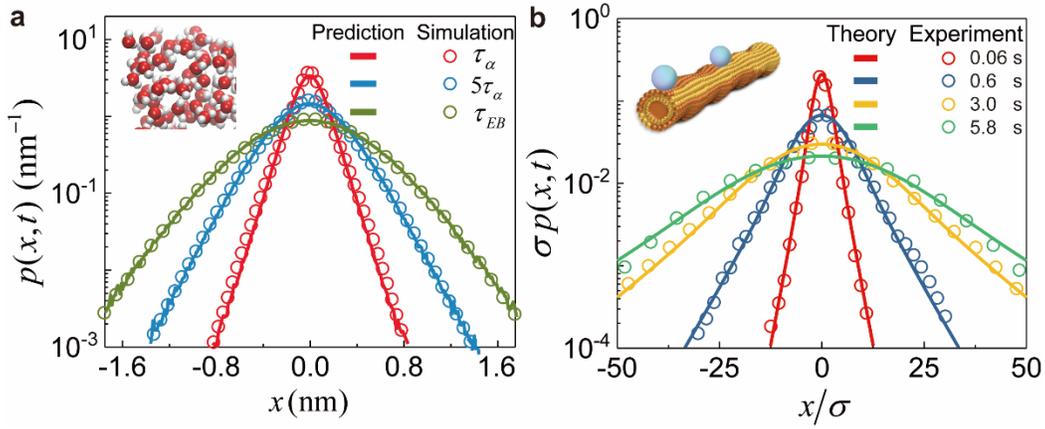

**FIG. 3 | Quantitative explanation of displacement distributions for the supercooled water and colloidal beads on lipid tubes. a,** Displacement distributions, $p(x,t)$, along $x$ axis at three different times, $\tau_\alpha (\cong 2.9 \text{ ns})$, $5\tau_\alpha (\cong 14.4 \text{ ns})$, and $\tau_{EB} (\cong 36.4 \text{ ns})$. (circles) Simulation results for the TIP4P/2005 water system at 193 K, (lines) theoretical predictions of our first model, $D_\Gamma(t) = A\exp(-\beta E_\Gamma(t))$ with $E_\Gamma(t) = \langle E \rangle + \sum_{k=1}^{3} b_k \Gamma_k(t)$, where $\{b_k\}$ and $\{\Gamma_k(t)\}$ are constants and stationary Gaussian Markov processes, also known as Ornstein-Uhlenbeck (OU) processes, respectively. **b,** Scaled displacement distributions, $\sigma p(x,t)$, along the $x$ axis at various times for colloidal beads with diameter $\sigma$ moving on lipid tubes at various times[36]: (circles) experimental results reported in ref. [36], (lines) theoretical results of our second model, $D_\Gamma = \langle D \rangle (a_1 \Gamma_1 + a_2 \Gamma_2)$, where $\{a_i\}$ and $\{\Gamma_k(t)\}$ are constants and OU processes. In both models, OU processes satisfy the following property: $\langle \Gamma_i(t) \Gamma_j(t) \rangle = \delta_{ij} \exp(-\lambda_i t)$ (see Methods). Although the relaxation of the non-Gaussian displacement distribution occurs in vastly different time scales between supercooled water and colloidal beads, the two systems show remarkably similar relaxation dynamics, demonstrating the universality in thermal motion dynamics of complex fluid systems.

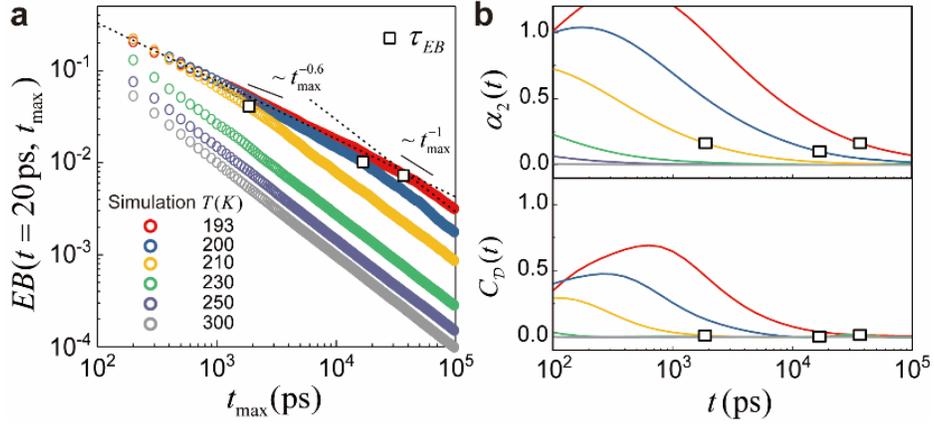

**FIG. 4 | Ergodicity breaking parameter, the non-Gaussian parameter, and the diffusion kernel correlation function of supercooled water. a,** Dependence of the EB parameter on the maximum length, $t_{max}$, of displacement time traces when the lag time, $t$, is fixed to 20 ps for the TIP4P/2005 water system. At $T = 230$ K or higher, the EB paramter behaves as $EB(t, t_{max}) \sim t/t_{max}$, which is expected for the case where time traces are statistically homogeneous. However, at temperatures lower than 230 K, the EB parameter shows anomalous power-law behavior with exponents smaller than $-1$ at short times and reattains homogeneous behavior at long times. $\tau_{EB}$ indicates the time at which the crossover between the two regimes occurs. **b,** The NGP, $\alpha_2(t)$, and the diffusion kernel correlation function, $C_\mathcal{D}(t)$, at various temperatures. At $\tau_{EB}$, the NGP remains substantially finite while the diffusion kernel correlation function fully relaxes to zero. This shows that only after the long-time relaxation of $C_\mathcal{D}(t)$ is complete, does the EB parameter resume the time dependence expected for homogeneous systems.

| Supercooled water | | Value |
|---|---|---|
| Mean square displacement [in equation (5)] | $c_0, c_1, c_2$ | $4.03\times10^{-5}, 6.28\times10^{-4}, 9.99\times10^{-1}$ |
| | $\gamma_0, \gamma_1, \gamma_2$ | $1.16$ ps$^{-1}$, $6.14$ ps$^{-1}$, $11.9$ ps$^{-1}$ |
| | $\omega_1, \omega_2$ | $6.13$ ps$^{-1}$, $2.46\times10^{-8}$ ps$^{-1}$ |
| Time correlation function of diffusion coefficient, $\phi_D(t)$ [in equations (7) and (49c)] | $\beta b_1, \beta b_2, \beta b_3$ | $1.51\times10^{-2}, 2.12\times10^{-1}, 6.19\times10^{-1}$ |
| | $\lambda_1, \lambda_2, \lambda_3$ | $7.61\times10^{-3}$ ns$^{-1}$, $1.30\times10^{-1}$ ns$^{-1}$, $5.84\times10^{-1}$ ns$^{-1}$ |
| $\langle\delta c_i(t)\delta c_j(0)\rangle$ $[=\zeta_{ij}\exp(-\lambda_{ij}t)]$ [in equation (7)] | $\zeta_{11}, \zeta_{22}$ | $3.48\times10^{-2}, 2.56$ |
| | $\zeta_{01}, \zeta_{02}, \zeta_{12}$ | $1.14\times10^{-7}, -7.43\times10^{-5}, -2.98\times10^{-1}$ |
| | $\lambda_{11}, \lambda_{22}$ | $4.88\times10^{2}$ ps$^{-1}$, $3.70\times10^{1}$ ps$^{-1}$ |
| | $\lambda_{01}, \lambda_{10}$ | $2.14\times10^{1}$ ps$^{-1}$, $2.90\times10^{-3}$ ps$^{-1}$ † |
| | $\lambda_{02}, \lambda_{20}$ | $2.51\times10^{2}$ ps$^{-1}$, $1.72\times10^{-2}$ ps$^{-1}$ † |
| | $\lambda_{12}, \lambda_{21}$ | $1.45\times10^{1}$ ps$^{-1}$, $1.45\times10^{1}$ ps$^{-1}$ |

**Table 1. Optimized values of adjustable parameters for supercooled water self-diffusion.** The MSD is analyzed by equation (5) with $n=2$. The diffusion kernel correlation $C_D(t)$ is analyzed by the inverse Laplace transform of equation (7). We use the diffusion coefficient fluctuation model described above equation (47) and use equation (49c) for the TCF, $\phi_D(t)$, of the diffusion coefficient fluctuation, which appears in the first term on the R.H.S. of equation (7). Because $\phi_D(t)$ governs the long-time relaxation of $C_D(t)$, $\phi_D(t)$ can be separately optimized against the time profile of $C_D(t)$ at times longer than the NGP peak time (see Supplementary Fig. 10). $\{\zeta_{ij}\}$ and $\{\lambda_{ij}\}$ are parameters determining the weight coefficient correlations, $\{\langle\delta c_i(t)\delta c_j(0)\rangle\}$, which appear in the second term on the R.H.S. of equation (7). The parameter values are optimized against the entire $C_D(t)$ time profile data of supercooled water at 193 K, shown in Fig. 2 (see Methods). †Any one of the two tabulated values can be assigned to either $\lambda_{ij}$ or $\lambda_{ji}$. This can be seen by noting that the second term on the R.H.S. of equation (7) can be rewritten as

$$\sum_{i=1}^{n}\frac{\zeta_{ii}}{s+\lambda_{ii}}\hat{f}_i^2(s)+\sum_{i=0}^{n}\sum_{j=i+1}^{n}\zeta_{ij}\left[\frac{1}{(s+\lambda_{ij})}+\frac{1}{(s+\lambda_{ji})}\right]\hat{f}_i(s)\hat{f}_j(s),$$

which is invariant under the exchange of $\lambda_{ij}$ and $\lambda_{ji}$.

| Colloidal beads on lipid tubes | | Value |
|---|---|---|
| $p(x,t)$ | $a_1, a_2$ | $4.02\times10^{-1}, 5.98\times10^{-1}$ |
| [in equation (53)] | $\lambda_1, \lambda_2$ | $3.43\text{ s}^{-1}, 1.04\times10^{-2}\text{ s}^{-1}$ |

**Table 2. Optimized values of adjustable parameters for colloidal bead diffusion on lipid tubes.** The experimental data for the time-dependent displacement distribution of colloidal beads reported in ref. [36] are analyzed by equation (53) (see Fig. 3b). For the model of the diffusion coefficient fluctuation defined in equation (51), $\{a_i\}$ and $\{\lambda_i\}$ are related to the relative variance, the time correlation function of the diffusion coefficient fluctuation, and the displacement distribution by equations (52a), (52b), and (53), respectively. This system exhibits Fickian diffusion in the experimental time scale. The diffusion constant is estimated to be $\langle D \rangle/\sigma^2 \cong 41.5\text{ Hz}$, where $\sigma$ denotes the diameter of a colloidal bead. Using the optimized parameter values, we can also calculate the time profiles of the NGP and $C_\mathcal{D}(t)$ using equation (45) and (52b) (see Supplementary Fig. 5).

## Methods

**Derivation of equation (1).** Let $p(m, \Gamma, t)$ denote the joint probability that a random walker is located at the $m$th site and the environmental state is at $\Gamma$ at time $t$. $p(m, \Gamma, t)$ can then be written as[45, 58]

$$p(m, \Gamma, t) = \sum_{N=0}^{\infty} p(m|N) P_N(\Gamma, t) \tag{8}$$

where $p(m|N)$ is given by $p(m|N) = (N!/m_+!m_-!) 2^{-N}$ with $m_\pm = (N \pm m)/2$ when $N \geq |m|$, and $p(m|N) = 0$ when $N < |m|$. In equation (8), $P_N(\Gamma, t)$ denotes the joint probability that the total number of jumps made by a random walker is $N$ and the environmental state is at $\Gamma$ at time $t$. $P_N(\Gamma, t)$ satisfies Sung and Silbey's generalized master equation[59]:

$$\hat{\dot{P}}_N(\Gamma, s) = \hat{\kappa}_\Gamma(s)\left[\hat{P}_{N-1}(\Gamma, s) - \hat{P}_N(\Gamma, s)\right] + L(\Gamma)\hat{P}_N(\Gamma, s) \tag{9}$$

Substituting equation (9) into the Laplace transform of the time derivative of equation (8), we obtain

$$\hat{\dot{p}}(m, \Gamma, s) = \sum_{N=0}^{\infty} p(m|N) \left\{ \hat{\kappa}_\Gamma(s)[\hat{P}_{N-1}(\Gamma, s) - \hat{P}_N(\Gamma, s)] + L(\Gamma)\hat{P}_N(\Gamma, s) \right\} \tag{10}$$

By noting that $\sum_{N=0}^{\infty} p(m|N) \hat{P}_{N-1}(\Gamma, s) = \sum_{N=-1}^{\infty} p(m|N+1) \hat{P}_N(\Gamma, s)$ and $\hat{p}_{-1}(\Gamma, s) = 0$, we can rewrite equation (10) as

$$\hat{\dot{p}}(m, \Gamma, s) = \sum_{N=0}^{\infty} \hat{\kappa}_\Gamma(s)\left[p(m|N+1) - p(m|N)\right] \hat{P}_N(\Gamma, s) + L(\Gamma)\hat{p}(m, \Gamma, s) \tag{11}$$

It is easy to show that $p(m|N+1)$ satisfies

$$p(m|N+1) = \frac{1}{2}[p(m-1|N) + p(m+1|N)] \tag{12}$$

The substitution of equation (12) into equation (11) yields

$$\hat{p}(m,\Gamma,s) = \frac{\hat{\kappa}_\Gamma(s)}{2}[\hat{p}(m-1,\Gamma,s) + \hat{p}(m+1,\Gamma,s) - 2\hat{p}(m,\Gamma,s)] + L(\Gamma)\hat{p}(m,\Gamma,s) \tag{13}$$

In terms of position, $x(=m\varepsilon)$ with $\varepsilon$ being a lattice constant, equation (13) can be written as

$$\hat{p}(x,\Gamma,s) = \frac{\hat{\kappa}_\Gamma(s)}{2}[\hat{p}(x-\varepsilon,\Gamma,s) + \hat{p}(x+\varepsilon,\Gamma,s) - 2\hat{p}(x,\Gamma,s)] + L(\Gamma)\hat{p}(m,\Gamma,s) \tag{14}$$

where $\hat{p}(x,\Gamma,s)$ denotes the Laplace transform of the joint probability density, $p(x,\Gamma,t)$, of the random walker position, $x$, and the environmental state, $\Gamma$, at time $t$. Using the Taylor expansions of $\hat{p}(x\pm\varepsilon,\Gamma,s)$ with respect to $\varepsilon$ and taking the small $\varepsilon$ limit, i.e., $\hat{p}(x\pm\varepsilon,\Gamma,s) \cong \hat{p}(x,\Gamma,s) \pm \partial_x\hat{p}(x,\Gamma,s)\varepsilon + \frac{1}{2}\partial_x^2\hat{p}(x,\Gamma,s)\varepsilon^2$, we obtain the following transport equation from equation (14):

$$\hat{p}(x,\Gamma,s) = \hat{\mathcal{D}}_\Gamma(s)\frac{\partial^2}{\partial x^2}\hat{p}(\mathbf{r},\Gamma,s) + L(\Gamma)\hat{p}(x,\Gamma,s) \tag{15}$$

where $\hat{\mathcal{D}}_\Gamma(s)$ is defined by $\hat{\mathcal{D}}_\Gamma(s) = \varepsilon^2\hat{\kappa}_\Gamma(s)/2$. $\hat{\mathcal{D}}_\Gamma(s)$ denotes the environment state-dependent diffusion kernel.

The $d$-dimensional extension of equation (15) is simply given by

$$\hat{p}(\mathbf{r},\Gamma,s) = \hat{\mathcal{D}}_\Gamma(s)\nabla^2 \hat{p}(\mathbf{r},\Gamma,s) + L(\Gamma)\hat{p}(\mathbf{r},\Gamma,s) \tag{16}$$

with $\hat{\mathcal{D}}_\Gamma(s) = \varepsilon^2 \hat{\kappa}_\Gamma(s)/2d$ and **r** being a position vector of a tracer particle in $d$-dimensional space. Equation (16) is equivalent to equation (1) in the main text.

**Derivation of equations (2a) and (2b).** In equation (1), $\hat{\dot{p}}(\mathbf{r},\Gamma,s)$ can be written as $\hat{\dot{p}}(\mathbf{r},\Gamma,s) = s\hat{p}(\mathbf{r},\Gamma,s) - p(\mathbf{r},\Gamma,0)$, where $p(\mathbf{r},\Gamma,0)$ is given here by $p(\mathbf{r},\Gamma,0) = \delta(\mathbf{r}-\mathbf{r}_0)p_{st}(\Gamma)$ with $p_{st}(\Gamma)$ denoting the stationary distribution of the environmental state, $\Gamma$. Performing the Fourier transform of equation (1) with $\hat{\dot{p}}(\mathbf{r},\Gamma,s) = s\hat{p}(\mathbf{r},\Gamma,s) - \delta(\mathbf{r}-\mathbf{r}_0)p_{st}(\Gamma)$ over **r**, we obtain

$$s\hat{\tilde{p}}(\mathbf{k},\Gamma,s) - e^{i\mathbf{k}\cdot\mathbf{r}_0}p_{st}(\Gamma) = -k^2\hat{\mathcal{D}}_\Gamma(s)\hat{\tilde{p}}(\mathbf{k},\Gamma,s) + L(\Gamma)\hat{\tilde{p}}(\mathbf{k},\Gamma,s) \tag{17}$$

where $\hat{\tilde{p}}(\mathbf{k},\Gamma,s)$ and $k$ respectively denote the Fourier transform of $\hat{p}(\mathbf{r},\Gamma,s)$ defined by $\hat{\tilde{p}}(\mathbf{k},\Gamma,s) = \int d\mathbf{r}\, e^{i\mathbf{k}\cdot\mathbf{r}} \hat{p}(\mathbf{r},\Gamma,s)$ and the magnitude of the wave vector, **k**, i.e. $k = |\mathbf{k}|$. By setting the initial position, $\mathbf{r}_0$, to be the origin of the coordinate system, explicitly, $\mathbf{r}_0 = \mathbf{0}$, **r** can be then regarded as a displacement vector.

The first two non-vanishing moments, $\langle |\mathbf{r}(t)|^2\rangle \left(=\langle r^2(t)\rangle \equiv \Delta_2(t)\right)$ and $\langle |\mathbf{r}(t)|^4\rangle \left(=\langle r^4(t)\rangle \equiv \Delta_4(t)\right)$, are related to the $\Gamma$-dependent displacement distribution, $p(\mathbf{r},\Gamma,t)$, as

$$\Delta_{q=2,4}(t) = \int_0^\infty dr\, \gamma_d r^{d-1} r^q p(\mathbf{r},\Gamma,t) \tag{18}$$

with $\gamma_d$ denoting the dimension-dependent factor given by $\gamma_d = 2\pi^{d/2}/\Gamma(d/2)$. $\gamma_d r^{d-1}$ is the surface area of a $d$-dimensional sphere with radius $r$. Taking the second and fourth derivatives of $\tilde{p}(\mathbf{k}, \Gamma, t)$ with respect to $k$ and setting $k = 0$ in the resulting equations, we have

$$\begin{aligned}
\partial_k^q \tilde{p}(\mathbf{k}, \Gamma, t)\Big|_{k=0} &= \partial_k^q \int d\mathbf{r} e^{ikr\cos\theta} p(\mathbf{r}, \Gamma, t)\Big|_{k=0} \\
&= i^q \int d\mathbf{r}\left[(r\cos\theta)^q p(\mathbf{r}, \Gamma, t)\right] \\
&= i^q \int_0^\infty dr \int_0^\pi d\theta \gamma_{d-1} r^{d-1} \sin^{d-2}\theta \left[r^q \cos^q\theta p(\mathbf{r}, \Gamma, t)\right] \\
&= \begin{cases} -\Delta_2(t)/d, & \text{for } q = 2 \\ 3\Delta_4(t)/d(2+d), & \text{for } q = 4 \end{cases}
\end{aligned} \quad (19)$$

where $\theta$ is the angle between the two vectors, $\mathbf{k}$ and $\mathbf{r}$, defined by $\cos\theta = \mathbf{k}\cdot\mathbf{r}/kr$. In equation (19), the third equality holds for the $d$-dimensional spherical coordinate system. For the fourth equality, we have used the following equation: $\gamma_{d-1}\int_0^\pi d\theta \sin^{d-2}\theta \cos^q\theta = 2\pi^{\frac{d-1}{2}}\Gamma(\frac{1+q}{2})/\Gamma(\frac{d+q}{2})$ when $q$ is even. Applying the operation, $-d\partial_k^2(\cdots)\Big|_{k=0}$, to both sides of equation (17) and using equation (19), we obtain

$$s\hat{\Delta}_2(\Gamma, s) = 2d\hat{\mathcal{D}}_\Gamma(s)\hat{\tilde{p}}(0, \Gamma, s) + L(\Gamma)\hat{\Delta}_2(\Gamma, s) \quad (20)$$

With $\hat{\tilde{p}}(0, \Gamma, s) = \int d\mathbf{r}\hat{p}(\mathbf{r}, \Gamma, s) = p_{st}(\Gamma)/s$ at hand, we can obtain the expression of $\hat{\Delta}_2(\Gamma, s)$ from equation (20), which is given by

$$\hat{\Delta}_2(\Gamma, s) = 2d[s - L(\Gamma)]^{-1}\frac{\hat{\mathcal{D}}_\Gamma(s)p_{st}(\Gamma)}{s} \quad (21)$$

Using the property of the Dirac delta function and $\hat{G}(\Gamma,s|\Gamma_0) = [s - L(\Gamma)]^{-1}\delta(\Gamma - \Gamma_0)$, where $\hat{G}(\Gamma,s|\Gamma_0)$ denotes the Laplace transform of the propagator, $G(\Gamma,t|\Gamma_0)$, or the conditional probability density that the environment is at state $\Gamma$ at time $t$, given that the environment was at state $\Gamma_0$ at time 0, equation (21) can be rewritten as

$$\hat{\Delta}_2(\Gamma,s) = 2d\int d\Gamma_0 [s - L(\Gamma)]^{-1}\delta(\Gamma - \Gamma_0)\frac{\hat{\mathcal{D}}_{\Gamma_0}(s)p_{st}(\Gamma_0)}{s}$$
$$= 2d\int d\Gamma_0 \hat{G}(\Gamma,s|\Gamma_0)\frac{\hat{\mathcal{D}}_{\Gamma_0}(s)p_{st}(\Gamma_0)}{s} \quad (22)$$

Integrating both sides of equation (22) over $\Gamma$ and using the normalization condition, $\int d\Gamma \hat{G}(\Gamma,s|\Gamma_0) = 1/s$, we finally obtain the expression of the mean square displacement, $\hat{\Delta}_2(s)\left[=\int d\Gamma \hat{\Delta}_2(\Gamma,s)\right]$, in the Laplace domain:

$$\hat{\Delta}_2(s) = \frac{2d}{s^2}\langle \hat{\mathcal{D}}_\Gamma(s)\rangle \quad (23)$$

where $\langle \hat{\mathcal{D}}_\Gamma(s)\rangle$ denotes the averaged diffusion kernel defined by $\langle \hat{\mathcal{D}}_\Gamma(s)\rangle = \int d\Gamma \hat{\mathcal{D}}_\Gamma(s) p_{st}(\Gamma)$.

Applying the operation, $d(2+d)\partial_k^4(\cdots)\big|_{k=0}/3$, to both sides of equation (17) and using equation (19), we obtain

$$s\hat{\Delta}_4(\Gamma,s) = \frac{8d(d+2)}{s}\hat{\mathcal{D}}_\Gamma(s)[s - L(\Gamma)]^{-1}\hat{\mathcal{D}}_\Gamma(s)p_{st}(\Gamma) + L(\Gamma)\hat{\Delta}_4(\Gamma,s) \quad (24)$$

From equation (24), we have

$$\hat{\Delta}_4(\Gamma, s) = \frac{8d(d+2)}{s}[s-L(\Gamma)]^{-1}\hat{\mathcal{D}}_\Gamma(s)[s-L(\Gamma)]^{-1}\hat{\mathcal{D}}_\Gamma(s)p_{st}(\Gamma)$$
$$= \frac{8d(d+2)}{s}\int d\Gamma_1 \int d\Gamma_0 \hat{G}(\Gamma, s|\Gamma_1)\hat{\mathcal{D}}_{\Gamma_1}(s)\hat{G}(\Gamma_1, s|\Gamma_0)\hat{\mathcal{D}}_{\Gamma_0}(s)p_{st}(\Gamma_0) \quad (25)$$

whose integration over $\Gamma$ yields the expression of $\hat{\Delta}_4(s)\left[=\int d\Gamma \hat{\Delta}_4(\Gamma, s)\right]$:

$$\hat{\Delta}_4(s) = \frac{8d(d+2)}{s^2}\int d\Gamma \int d\Gamma_0 \hat{\mathcal{D}}_\Gamma(s)\hat{G}(\Gamma, s|\Gamma_0)\hat{\mathcal{D}}_{\Gamma_0}(s)p_{st}(\Gamma_0) \quad (26)$$

Equation (26) can be rearranged to

$$\hat{\Delta}_4(s) = \frac{8d(d+2)}{s^3}\langle \hat{\mathcal{D}}_\Gamma(s)\rangle^2 \left[1+s\hat{C}_\mathcal{D}(s)\right] \quad (27)$$

where $\hat{C}_\mathcal{D}(s)$ denotes the Laplace transform of the generalized time correlation function, defined by

$$\hat{C}_\mathcal{D}(s) = \int d\Gamma \int d\Gamma_0 \frac{\delta\hat{\mathcal{D}}_\Gamma(s)}{\langle \hat{\mathcal{D}}_\Gamma(s)\rangle}\hat{G}(\Gamma, s|\Gamma_0)\frac{\delta\hat{\mathcal{D}}_{\Gamma_0}(s)}{\langle \hat{\mathcal{D}}_{\Gamma_0}(s)\rangle}p_{st}(\Gamma_0) \quad (28)$$

In equation (26), $\delta\hat{\mathcal{D}}_\Gamma(s)$ denotes the deviation, $\delta\hat{\mathcal{D}}_\Gamma(s) = \hat{\mathcal{D}}_\Gamma(s) - \langle \hat{\mathcal{D}}_\Gamma(s)\rangle$, of the diffusion kernel from its average. Substituting $\langle \hat{\mathcal{D}}_\Gamma(s)\rangle = s^2\hat{\Delta}_2(s)/2d$, or equation (23), into equation (27), we finally obtain

$$\hat{\Delta}_4(s) = \left(1+\frac{2}{d}\right)2s\hat{\Delta}_2(s)^2\left[1+s\hat{C}_\mathcal{D}(s)\right] \quad (29)$$

Equations (23) and (29) are equivalent to equations (2a) and (2b) in the main text.

**Time correlation function of squared speed.** Here, we provide a short derivation of the time correlation function of squared speed, $v^2$, which describes the short-time behavior of the diffusion kernel correlation function, $C_D(t)$, discussed in the main text. Let $v_\alpha$ denote the $\alpha$ component of a velocity vector, **v**, which follows the Maxwell-Boltzmann distribution. The time correlation function of $v_\alpha^2$, $\langle v_\alpha^2(t) v_\alpha^2(0) \rangle$ can be regarded as the fourth-order correlation, $\langle v_\alpha(t) v_\alpha(t) v_\alpha(0) v_\alpha(0) \rangle$, which can be easily calculated as

$$\langle v_\alpha(t) v_\alpha(t) v_\alpha(0) v_\alpha(0) \rangle = \langle v_\alpha(t) v_\alpha(t) \rangle \langle v_\alpha(0) v_\alpha(0) \rangle + 2 \langle v_\alpha(t) v_\alpha(0) \rangle \langle v_\alpha(t) v_\alpha(0) \rangle \tag{30}$$

because $v_\alpha$ is a Gaussian random variable at thermal equilibrium. After rearranging equation (30), we have

$$\begin{aligned} \langle \delta v_\alpha^2(t) \delta v_\alpha^2(0) \rangle &= \langle v_\alpha^2(t) v_\alpha^2(0) \rangle - \langle v_\alpha^2 \rangle^2 \\ &= 2 \langle v_\alpha(t) v_\alpha(0) \rangle^2 \end{aligned} \tag{31}$$

Dividing both sides of equation (31) by $\langle v_\alpha^2 \rangle^2$, we obtain the mean-scaled correlation function of $v_\alpha^2$, which is given by

$$\frac{\langle \delta v_\alpha^2(t) \delta v_\alpha^2(0) \rangle}{\langle v_\alpha^2 \rangle^2} = 2 \phi_{v_\alpha}(t)^2 \tag{32}$$

where $\phi_{v_\alpha}(t) \left[ = \langle v_\alpha(t) v_\alpha(0) \rangle / \langle v_\alpha^2 \rangle \right]$ denotes the normalized time correlation function of $v_\alpha$. Noting the well-known relation, $\Delta_2(t) = 2 \int_0^t d\tau_2 \int_0^{\tau_2} d\tau_1 \langle \mathbf{v}(\tau_2 - \tau_1) \cdot \mathbf{v}(0) \rangle$, $\phi_{v_\alpha}(t)$ can be calculated as the second-order time derivative of the MSD, explicitly, $\phi_{v_\alpha}(t) = \ddot{\Delta}_2(t)/2d\langle v_\alpha^2 \rangle$ with $\langle v_\alpha^2 \rangle = k_B T/M$. In terms of squared speed, $v^2 \left(= \sum_{\alpha=x,y,z} v_\alpha^2\right)$, equation (32) can be rewritten as

$$\frac{\langle \delta v_\alpha^2(t) \delta v_\alpha^2(0) \rangle}{\langle v_\alpha^2 \rangle^2} = d \frac{\langle \delta v^2(t) \delta v^2(0) \rangle}{\langle v^2 \rangle^2} \tag{33}$$

by noting that $\langle v^2 \rangle$ and $\langle v^2(t) v^2(0) \rangle$ are related to $\langle v_\alpha^2 \rangle$ and $\langle v_\alpha^2(t) v_\alpha^2(0) \rangle$ as $\langle v^2 \rangle = d \langle v_\alpha^2 \rangle$ and $\langle \delta v^2(t) \delta v^2(0) \rangle = d \langle \delta v_\alpha^2(t) \delta v_\alpha^2(0) \rangle$. Equation (32) or (33) gives the short-time profile of $C_D(t)$ as shown in Fig. 2d and Supplementary Fig. 1d (see also Supplementary Fig. 8).

**Long-time asymptotic behavior of the non-Gaussian parameter.** In this method section, we derive equation (3b) in the main text. At long times, the MSD assumes $\Delta_2(t) \cong 2d\bar{D}t + \Delta_c$ [see equation (3a)], whose Laplace transform is given by

$$\hat{\Delta}_2(s) \cong \frac{2d\bar{D}}{s^2} + \frac{\Delta_c}{s} \tag{34}$$

Substituting equation (34) into equation (29), we obtain

$$\hat{\Delta}_4(s) \cong 2\left(1 + \frac{2}{d}\right)\left(\frac{4d^2\bar{D}^2}{s^3} + \frac{4d\bar{D}\Delta_c}{s^2} + \frac{\Delta_c^2}{s}\right)\left[1 + s\hat{C}_D(s)\right] \tag{35}$$

The inverse Laplace transform of equation (35) is given by

$$\Delta_4(t) \cong 2\left(1 + \frac{2}{d}\right)\left[2d^2\bar{D}^2 t^2 + 4d\bar{D}\Delta_c t + \Delta_c^2 + \right.$$
$$\left. + 4d^2\bar{D}^2 \int_0^t dt'(t-t')C_D(t') + 4d\bar{D}\Delta_c \int_0^t dt' C_D(t') + \Delta_c^2 C_D(t)\right] \tag{36}$$

At long times, equation (36) further reduces to

$$\Delta_4(t) \cong 2\left(1+\frac{2}{d}\right)\left[2d^2\bar{D}^2t^2 + \left(4d\bar{D}\Delta_c + 4d^2\bar{D}^2\hat{C}_\mathcal{D}(0)\right)t + \Delta_c^2 + 4d\bar{D}\Delta_c\hat{C}_\mathcal{D}(0)\right] \qquad (37)$$

Substituting equations (3a) and (37) into the definition of the NGP, $\alpha_2(t)\left[=\Delta_4(t)/\left[(1+2/d)\Delta_2(t)^2\right]-1\right]$, we obtain the long-time expression of the NGP as follows:

$$\begin{aligned}\alpha_2(t) &= \frac{\Delta_4(t)}{(1+2/d)\Delta_2(t)^2} - 1 \\ &\cong 2\frac{2d^2\bar{D}^2t^2 + \left(4d\bar{D}\Delta_c + 4d^2\bar{D}^2\hat{C}_\mathcal{D}(0)\right)t + \Delta_c^2 + 4d\bar{D}\Delta_c\hat{C}_\mathcal{D}(0)}{4d^2\bar{D}^2t^2 + 4d\bar{D}\Delta_c t + \Delta_c^2} - 1\end{aligned} \qquad (38)$$

Taking only the leading-order term on the right-hand side of equation (36), we obtain equation (3b) in the main text. This conclusion is valid only when $C_\mathcal{D}(\infty)=0$. When $C_\mathcal{D}(\infty)$ is a nonzero value, equation (36) can be rewritten as $\Delta_4(t) \cong 2(1+2/d)\left(2d^2\bar{D}^2t^2 + 4d\bar{D}\Delta_c t + \Delta_c^2\right)[1+C_\mathcal{D}(\infty)]$ by noting that $\lim_{s\to 0} s\hat{C}_\mathcal{D}(s) = C_\mathcal{D}(\infty)$ in equation (35). With this equation, we reach the following conclusion from the definition of the NGP: $\alpha_2(\infty) = C_\mathcal{D}(\infty)$.

**Extraction of diffusion kernel correlation function from MSD and NGP.** Here, we present the procedure for extracting the diffusion kernel correlation function, $C_\mathcal{D}(t)$, from the MSD and NGP obtained by computer simulation, where $C_\mathcal{D}(t)$ is defined by equation (28) in the Laplace domain. From equation (27), we can represent $\hat{C}_\mathcal{D}(s)$ in terms of the first two non-vanishing moments as follows:

$$\hat{C}_{\mathcal{D}}(s) = \frac{\hat{\Delta}_4(s)}{\left(1+\frac{2}{d}\right)2s^2\hat{\Delta}_2(s)^2} - \frac{1}{s} \tag{39}$$

On the other hand, the NGP, $\alpha_2(t)$, is defined by

$$\alpha_2(t) = \frac{\Delta_4(t)}{\left(1+\frac{2}{d}\right)\Delta_2(t)^2} - 1 \tag{40}$$

which can be rearranged with respect to the fourth moment, $\Delta_4(t)$, as

$$\Delta_4(t) = \left(1+\frac{2}{d}\right)\Delta_2(t)^2\left[1+\alpha_2(t)\right] \tag{41}$$

The simulation results for the MSD and NGP are well represented by equation (5) with two bound modes ($n = 2$) and a linear combination of three or four Gaussian shaped functions given by

$$\alpha_2(t) \cong \sum_{i=1}^{3 \text{ or } 4} a_i \exp\left[-(\log_{10} t - b_i)^2/c_i\right] \tag{42}$$

respectively. We perform the best fits of equations (5) and (42) to the simulation results for the MSD and NGP. By substituting the optimized results into equation (41), we obtain the analytic expression of $\Delta_4(t)$ as a function of time. Taking the Laplace transforms of the best fitted $\Delta_2(t)$ and $\Delta_4(t)$, and substituting the results into equation (39), we obtain the Laplace transform of $C_{\mathcal{D}}(t)$ for a given set of the MSD and NGP data. To obtain the value of $C_{\mathcal{D}}(t)$ at a given time, $t$, we perform the numerical Laplace inversion of equation (39) using the Stehfest algorithm[60].

**NGP for stochastic diffusivity model.** Here, we derive the NGP of the SD model from equations (2a) and (2b). The MSD and the fourth moment for the SD model can be regarded as the asymptotic expressions of equations (2a) and (2b) in the small-$s$ limit, explicitly,

$$\hat{\Delta}_2(s) \cong \frac{2d\bar{D}}{s^2} \tag{43a}$$

$$\hat{\Delta}_4(s) \cong \left(1+\frac{2}{d}\right)\frac{8d^2\bar{D}^2}{s^3}\left[1+s\eta_D^2\hat{\phi}_D(s)\right] \tag{43b}$$

where $\bar{D}$ denotes the small-$s$ limit of the averaged diffusion kernel, $\langle\hat{\mathcal{D}}_\Gamma(s)\rangle$, in the Laplace domain. Equation (43b) is obtained by substituting equation (43a) and the small-$s$ limit expression, $\eta_D^2\hat{\phi}_D(s)$, of $\hat{C}_D(s)$ into equation (2b). The inverse Laplace transformations of equations (43a) and (43b) respectively give

$$\Delta_2(t) \cong 2d\bar{D}t \tag{44a}$$

$$\Delta_4(t) \cong 2\left(1+\frac{2}{d}\right)\left[2d^2\bar{D}^2t^2 + 4d^2\bar{D}^2\int_0^t dt'(t-t')\eta_D^2\phi_D(t')\right] \tag{44b}$$

Finally, the NGP for the SD model is obtained by substituting equations (44a) and (44b) into equation (40):

$$\alpha_2(t) = \frac{2\eta_D^2}{t^2}\int_0^t dt'(t-t')\phi_D(t')$$
$$= \begin{cases} \eta_D^2 & \text{as } t \to 0 \\ \dfrac{2\eta_D^2\hat{\phi}_D(0)}{t} & \text{as } t \to \infty \end{cases} \tag{45}$$

where the second equality holds in either the short-time or the long-time limit. When stochastic diffusivity is modeled as the square of the OU process, equation (45) reduces to the result given in ref. [45].

Equations (44a) and (45) suggest that, in the time scales where the MSD is linear in time, the initial NGP value is related to $\eta_D^2$. Since the MSD becomes linear in time only after the NGP peak time, it is reasonable to assume that the NGP value at the peak time, or the NGP peak height, is the same as the value of $\eta_D^2$. We find this is indeed case and the NGP peak value, $\alpha_2(\tau_{ng})$, is the same as the relative variance of diffusion coefficient, as demonstrated in Supplementary Fig. 4 for supercooled water. The second peak height, $C_\mathcal{D}(t^*)$, of $C_\mathcal{D}(t)$ is also related to the diffusion coefficient fluctuation (see Supplementary Fig. 9). In Fig. 3a and Supplementary Fig. 10, we use $\alpha_2(\tau_{ng})$ as an estimation of $\eta_D^2$.

**Analytic expression of diffusion kernel correlation (equation (7)).** Substituting equation (6) into the definition of $\hat{C}_\mathcal{D}(s)$ or equation (28), we obtain

$$\hat{C}_\mathcal{D}(s)\langle\hat{\mathcal{D}}_\Gamma(s)\rangle^2 = \hat{C}_{00}(s)\hat{f}_0(s)^2 + \sum_{i=0}^{n}\sum_{j=0}^{n}{}'\hat{C}_{ij}(s)\hat{f}_i(s)\hat{f}_j(s) \tag{46}$$

where $\hat{C}_{ij}(s)$ denotes the Laplace transform of the TCF of the weight coefficients, i.e., $\langle\delta c_i(t)\delta c_j(0)\rangle\left[\equiv \int d\Gamma \int d\Gamma_0 \delta c_i(\Gamma)G(\Gamma,t|\Gamma_0)\delta c_j(\Gamma_0)p_{st}(\Gamma_0)\right]$. The prime notation in the second term signifies that the sum does not include the term with $i=j=0$. Noting that $\hat{f}_0(s) \cong k_BT/M\gamma_0$ and $\hat{f}_{i\neq 0}(s) \cong 0$ in the small $s$ regime where $s \ll \gamma_i$, we obtain $\hat{\mathcal{D}}_\Gamma(0) = c_0(\Gamma)(k_BT/M\gamma_0)(\equiv D_\Gamma)$ from equation (6). Therefore, we obtain $C_{00}(t) = \langle \delta c_0(t) \delta c_0(0) \rangle = \langle \delta D(t) \delta D(0) \rangle (M\gamma_0/k_B T)^2$. We assume that, in equation (46), $C_{ij}(t)$, except for $C_{00}(t)$, can be described by $C_{ij}(t) = \langle \delta c_i(t) \delta c_j(0) \rangle = \zeta_{ij} \exp(-\lambda_{ij} t)$. $\{\zeta_{ij}\}$ should satisfy the following relation: $C_\mathcal{D}(0) = \sum_{i=0}^{2} \sum_{j=0}^{2} \zeta_{ij} = 2$ with $\zeta_{00}$ being identified as $\zeta_{00} = \langle c_0 \rangle^2 \eta_D^2$. This condition follows from the initial condition of the NGP, $\alpha_2(t=0) = 0$. To satisfy the initial condition of the NGP, the value of $C_\mathcal{D}(t=0) \left[ = \lim_{s \to 0} s \hat{C}_\mathcal{D}(s) \right]$ should be given by two (see Supplementary Note 4 in the Supplementary Information). Using this model, we obtain equation (7) in the main text. At long times where $t \gg \gamma_i^{-1}$, equation (46) reduces to $C_\mathcal{D}(t) \cong \langle \delta D(t) \delta D(0) \rangle / \langle D \rangle^2 = \phi_D(t) \eta_D^2$, consistent with the definition of $\hat{C}_\mathcal{D}(s)$ in the small $s$ limit where the diffusion kernel can be regarded as the diffusion coefficient, i.e., $\hat{\mathcal{D}}_\Gamma(s) \cong \hat{\mathcal{D}}_\Gamma(0) = D_\Gamma$. The explicit expression of $\langle \delta D(t) \delta D(0) \rangle$ is dependent on the model of $D_\Gamma$.

**First model of diffusion coefficient fluctuation.** When the diffusion coefficient depends on environmental variables through activation energy, i.e., $D_\Gamma = Ae^{-\beta E_\Gamma} = D_0 e^{-\beta \delta E_\Gamma}$ with $D_0$ being defined by $D_0 = Ae^{-\beta \langle E \rangle}$, the fluctuation, $\delta E_\Gamma$, in activation energy is modelled by a linear combination of $n_b$ number of independent OU processes $\Gamma_i(t)$:

$$\delta E_\Gamma = \sum_{k=1}^{n_b} b_k \Gamma_k \tag{47}$$

In equation (47), $b_k$ accounts for the relative contribution of the $k$th OU process, $\Gamma_k$, which has zero mean and unit variance, and obeys the following Langevin equation:

$$\partial_t \Gamma_k(t) = -\lambda_k \Gamma_k(t) + \xi_k(t) \tag{48}$$

where $\lambda_k$ and $\xi_k(t)$ respectively denote the relaxation rate of the TCF, $\langle \Gamma_k(t)\Gamma_k(0)\rangle (= e^{-t\lambda_k})$, and white Gaussian noise characterized by zero mean and $\langle \xi_k(t)\xi_k(t')\rangle = 2\lambda_k \delta(t-t')$. Because the statistics of $\Gamma_k$ follows a symmetric Gaussian distribution such that positive and negative values are equally probable, the sign of $b_k$ can be freely chosen. All $b_k$'s here are chosen to be positive.

Using the second-order cumulant expansion, $\langle e^z \rangle = e^{\langle z \rangle + \frac{1}{2}\langle \delta z^2 \rangle}$, which is exact when $z$ is a Gaussian random variable, we can calculate the mean, relative variance, and mean-scaled TCF of the diffusion coefficient as

$$\langle D \rangle = D_0 e^{\frac{1}{2}\beta^2 \langle \delta E^2 \rangle} \tag{49a}$$

$$\eta_D^2 = e^{\beta^2 \langle \delta E^2 \rangle} - 1 \tag{49b}$$

$$\eta_D^2 \phi_D(t) = \langle \delta D(t)\delta D(0)\rangle / \langle D \rangle^2 = e^{\beta^2 \langle \delta E^2 \rangle \phi_{\delta E}(t)} - 1 \tag{49c}$$

respectively. In equation (49c), the normalized TCF, $\phi_{\delta E}(t)$, of $\delta E_\Gamma$ is given by $\phi_{\delta E}(t) = \sum_k b_k^2 e^{-t\lambda_k} / \sum_k b_k^2 = \sum_k b_k' e^{-t\lambda_k}$ with $b_k'$ denoting $b_k' = b_k^2 / \sum_k b_k^2$. The values of $D_0$ and $\beta^2 \langle \delta E^2 \rangle (= \beta^2 \sum_k b_k^2)$ can be extracted by comparing equations (49a) and (49b) with the diffusion constant, $\bar{D}(= \lim_{t\to\infty} \Delta_2(t)/2dt)$, and the relative variance of the diffusion coefficient estimated from the NGP peak height, $\alpha_2(\tau_{ng})$, respectively (see text below equation (45) and Supplementary Fig. 4). The values of $\{b_k'\}$ and $\{\lambda_k\}$ characterizing $\phi_{\delta E}(t)$ can be obtained by comparing equation (48c) to the time profile of $C_\mathcal{D}(t)$ at times longer than the NGP peak time, $\tau_{ng}$ (see Supplementary Fig. 10). From the extracted values of $\beta^2\langle\delta E^2\rangle$ and $\{b'_k\}$, we can calculate the value of $\beta b_k$ by $\beta b_k = (\beta^2\langle\delta E^2\rangle b'_k)^{1/2}$.

Using the optimized model, we can predict the relaxation of the non-Gaussian displacement distribution in the Fickian diffusion regime at times longer than the NGP peak time. The theoretical prediction is found to be in excellent agreement with simulation results for the displacement distribution at various times. For this model, we obtain the displacement distribution using the following Brownian dynamics (BD) simulation algorithm for a tracer particle with a fluctuating diffusion coefficient; trajectories of one-dimensional displacement, $x$, are generated by

$$x(t+\Delta t) = x(t) + \sqrt{2D(t)\Delta t}\,\mathcal{N}[0,1] \tag{50a}$$

$$D(t) = D_0 \exp\left[-\sum_{k=1}^{n_b}(\beta b_k)\Gamma_k(t)\right] \tag{50b}$$

$$\Gamma_k(t+\Delta t) = \Gamma_k(t)e^{-\lambda_k\Delta t} + (1-e^{-2\lambda_k\Delta t})^{1/2}\mathcal{N}[0,1] \tag{50c}$$

where $\mathcal{N}[0,1]$ represents a standard Gaussian random variable with zero mean and unit variance. In the analysis shown in Figs. 2f and 3b, we consider three OU modes in equation (50b), i.e., $n_b = 3$.

One drawback of the numerical solution by the BD simulation is that it cannot accurately describe the non-Fickian transport dynamics emerging at times shorter than the NGP peak time, which affects the long-time dynamics of the MSD, NGP, and displacement distribution. For example, the MSD calculated by the BD simulation is directly proportional to the BD simulation time, $t'$, i.e., $\Delta_2^{BD}(t') = 2d\langle D\rangle t'$, which is inconsistent with the exact long-time MSD given in equation (3a). However, the MSD calculated by the BD simulation becomes consistent with equation (3a), valid at times longer than the NGP peak time, $\tau_{ng}$, if we identify the BD simulation time $t'$ as $t+t_s$ where $t_s$ is defined by $2d\langle D\rangle t_s = \Delta_c$, i.e., $\Delta_2(t) = \Delta_2^{BD}(t+t_s)$. Accordingly, the NGP and the displace distribution, $\alpha_{2,BD}(t')$ and $p_{BD}(x,t')$, calculated by the BD simulation are consistent with the exact long-time results, $\alpha_2(t)$ and $p(x,t)$, valid at times longer than the NGP peak time only when BD simulation time $t'$ is identified as $t+t_s$, i.e., $\alpha_2(t) = \alpha_{2,BD}(t+t_s)$ and $p(x,t) = p_{BD}(x,t+t_s)$. Using the BD simulation, we can obtain accurate results only at times longer than the NGP peak time, $\tau_{ng}$, where the diffusion kernel can be regarded as the diffusion coefficient, i.e., $\hat{\mathcal{D}}_\Gamma(s) \cong \hat{\mathcal{D}}_\Gamma(0) = D_\Gamma$. Since the diffusion coefficient has a physical meaning only after the NGP peak time and the value of the NGP peak height, $\alpha_2(\tau_{ng})$, is the same as the value of $\eta_D^2$, the initial value of the $\phi_D(t)\eta_D^2$, we assume that the dynamic fluctuation or time-dependent relaxation of the diffusion coefficient occurs only after the NGP peak time.

**Second Model of the diffusion coefficient fluctuation.** Here, we model the diffusion coefficient by $D_\Gamma = A_\Gamma e^{-\beta E}$, where $D_\Gamma$ is a linear combination of $n_a$ independent, squared OU processes. In this equation, the weight coefficients, $\{a_j\}$, are normalized by $\sum_j a_j = 1$. For this model, we have

$$D_\Gamma = \langle D\rangle \sum_{j=1}^{n_a} a_j \Gamma_j^2 \quad (a_j > 0) \tag{51}$$

where $\langle D \rangle = \langle A \rangle e^{-\beta E}$. The Langevin equation governing each OU process is given by equation (48). Using the fact that $\langle \Gamma_j^2(t)\Gamma_j^2(0)\rangle = \langle \Gamma_j^2 \rangle^2 + 2\langle \Gamma_j(t)\Gamma_j(0)\rangle^2 = 1 + 2e^{-2\lambda_j t}$, the relative variance and mean-scaled TCF of the diffusion coefficient can be calculated as

$$\eta_D^2 = 2\sum_{j=1}^{n_a} a_j^2 \tag{52a}$$

$$\eta_D^2 \phi_D(t) = 2\sum_{j=1}^{n_a} a_j^2 e^{-2\lambda_j t} \tag{52b}$$

Here, values of $\eta_D^2$ applicable to this model are $2/n_a \leq \eta_D^2 \leq 2$ in order for both conditions, $\sum_j a_j = 1$ and $\sum_j a_j^2 = \eta_D^2/2$, to be satisfied (see Supplementary Fig. 11). For the model described by equation (51), the analytical expression of the displacement distribution can be obtained as

$$p(x,t) = \frac{1}{2\pi}\int_{-\infty}^{\infty} dk e^{-ikx} \prod_{j=1}^{n_a}\left[\frac{4q_j e^{-(q_j-1)\lambda_j t}}{(q_j+1)^2 - (q_j-1)^2 e^{-2q_j \lambda_j t}}\right]^{1/2} \tag{53}$$

with $q_j$ denoting $q_j = (1 + 4a_j k^2 \langle D \rangle \lambda_j^{-1})^{1/2}$. Equation (53) is a simple extension of the previous result reported in refs. [44, 45], where $a_1 = a_2 = \cdots = a_{n_a}$ and $\lambda_1 = \lambda_2 = \cdots = \lambda_{n_a}$. Equation (53) can be used to determine the values of $\{a_j\}$ and $\{\lambda_j\}$ when the experimental data for the displacement distribution are available. For the colloidal bead diffusion data[36], the value of $\langle D \rangle$ is taken from the MSD shown in Supplementary Fig. 5a and the values of the other adjustable parameters are determined by fitting equation (53) to the multiple data sets at different times. Here, we use two modes, i.e., $n_a = 2$. In Fig. 3d, the NGP for the diffusion profile at 0.06 seconds is estimated to be roughly 0.75 so that the value of $\eta_D^2$ is expected to be similar to or larger than 0.75. Thus, it is necessary to use at least three modes, but the best fit of equation (53) with three modes varies only slightly from the result with two modes. The extracted values of the adjustable parameters are collected in Table 1.

**Quantitative Analysis of $C_D(t)$ of supercooled water at 193K in Fig. 2f.** For TIP4P/2005 supercooled water, we model $D_\Gamma$ as $D_\Gamma = Ae^{-\beta E_\Gamma} = D_0 e^{-\beta \delta E_\Gamma}$ with $E_\Gamma$ given by equation (47). For this model, the mean-scaled TCF, the analytic expression of $\eta_D^2 \phi_D(t) \left( = \langle \delta D(t) \delta D(0) \rangle / \langle D \rangle^2 \right)$, is given by equation (49c). As mentioned before, the values of adjustable parameters in equation (49c) can be obtained by fitting equation (49c) with three OU modes to the time profile of $C_D(t)$ at times longer than the NGP peak time, $\tau_{ng}$. The values of the other adjustable parameters in the second term on the R.H.S. of equation (7) are then obtained by fitting equation (7) to the time profile of $C_D(t)$ over the entire time range. The extracted values of the adjustable parameters are presented in Table 1.

**Simulation method for TIP4P/2005 water.** In Fig. 2, we performed molecular dynamics (MD) simulations in the NVT ensemble with 32000 TIP4P/2005 water molecules [55] in a cubic box and temperatures ranging from 193 K, near the hypothetical liquid-liquid critical point (LLCP) [56, 57], to 300 K along the isochore line at density, $\rho = 1.012$ g·cm$^{-3}$, close to the LLCP density [57]. At temperatures lower than the melting temperature, $T_m (\cong 250 \text{ K})$, of TIP4P/2005 water [56, 61], this system is in supercooled states. All MD simulations were performed using GROMACS 5.1.4 molecular dynamics simulation package [62]. In all cases, periodic boundary conditions were applied, and a time step of 2 fs with a Verlet integration[63] was used. The short-range interactions were truncated 9.5 Å. Long range electrostatic terms were computed using a particle mesh Ewald [64] with a grid spacing of 1.2 Å. Long range corrections were applied to the short-range Lennard-Jones interaction for both energy and pressure. Bond constraints were maintained using the LINCS (Linear Constraint Solver) algorithm [65]. To maintain constant temperature, we applied the Nosé-Hoover thermostat [66, 67] with the relaxation time given by 0.4 ps.

# Supplementary Information

# for

# Universal Transport Dynamics of Complex Fluids

**Supplementary Note 1 | Generalization of equation (1) for the complex fluids under a potential field.**

In this note, we present the generalization of equation (1) for the complex fluids under a potential field. Let $L(m)$ or $R(m)$ denote the probability that the random walker located at the $m$th site makes jump to the left or right adjacent site. Then the probability, $p(m|N)$, of finding the random walker at the $m$th site given that jump events occur $N$ times satisfies the following equation:

$$p(m|N+1) = R(m-1)p(m-1|N) + L(m+1)p(m+1|N) \tag{N1-1}$$

In the continuum limit, equation (N1-1) can be written as[1],

$$p(x|N+1) - p(x|N) = \frac{\varepsilon^2}{2}\frac{\partial}{\partial x}\left[\frac{\partial}{\partial x} + \beta\frac{\partial U(x)}{\partial x}\right]p(x|N) \tag{N1-2}$$

where $\varepsilon$ is a lattice constant, $\beta = 1/k_B T$, and $U(x)$ is the external potential given by

$$\beta\frac{\partial U}{\partial x} = \frac{2}{\varepsilon}[L(x) - R(x)].$$

Substituting the equation (N1-2) to equation (11) in Methods, we immediately obtain

$$\hat{p}(x,\Gamma,s) = \sum_{N=0}^{\infty} \frac{\hat{\kappa}_\Gamma(s)\varepsilon^2}{2}\frac{\partial}{\partial x}\left[\frac{\partial}{\partial x} + \beta\frac{\partial U(x)}{\partial x}\right]p(x|N)\hat{P}_N(\Gamma,s) + L(\Gamma)\hat{p}(x,\Gamma,s) \tag{N1-3}$$

Using the continuum limit version of equation (8), i.e., $p(x,\Gamma,t) = \sum_{N=0}^{\infty} p(x|N)P_N(\Gamma,t)$, we can obtain the following transport equation in the potential field,

$$\hat{p}(x,\Gamma,s) = \hat{\mathcal{D}}_\Gamma(s)\frac{\partial}{\partial x}\left[\frac{\partial}{\partial x} + \beta\frac{\partial U}{\partial x}\right]\hat{p}(x,\Gamma,s) + L(\Gamma)\hat{p}(x,\Gamma,s) \tag{N1-4}$$

where the definition of diffusion kernel, $\hat{\mathcal{D}}_\Gamma(s)$, is the same as that given below equation (1) in the main text. The $d$-dimensional extension of equation (N1-4) is simply given by

$$\hat{p}(\mathbf{r},\Gamma,s) = \hat{\mathcal{D}}_\Gamma(s)\nabla_\mathbf{r} \cdot [\nabla_\mathbf{r} + \beta\nabla_\mathbf{r} U(\mathbf{r})]\hat{p}(\mathbf{r},\Gamma,s) + L(\Gamma)\hat{p}(\mathbf{r},\Gamma,s) \qquad \text{(N1-5)}$$

where **r** denotes the position vector of a tracer particle in *d*-dimensional space.

We can further generalize equation (N1-5) for the transport system under a fluctuating potential as follows:

$$\hat{p}(\mathbf{r},\Gamma,s) = \hat{\mathcal{D}}_\Gamma(s)\nabla_\mathbf{r} \cdot [\nabla_\mathbf{r} + \beta\nabla_\mathbf{r} U(\mathbf{r},\Gamma)]\hat{p}(\mathbf{r},\Gamma,s) + L(\Gamma)\hat{p}(\mathbf{r},\Gamma,s) \qquad \text{(N1-6)}$$

where $U(\mathbf{r},\Gamma)$ denote the fluctuating potential dependent not only on the position of the tracer particle but also on hidden environmental variables. We leave applications of equations (N1-5) and (N1-6) for future researches.

**Supplementary Note 2 | Emergence of $\Delta_c$ from non-Poisson transport dynamics.**

In this note, we briefly discuss the emergence of finite $\Delta_c$ when the transport dynamics deviates from a simple Poisson process. Noting that $\hat{\kappa}_\Gamma(s) = s\hat{\psi}_\Gamma(s)/[1-\hat{\psi}_\Gamma(s)]$, one can obtain $\hat{\kappa}_\Gamma(0)$ and $\hat{\kappa}'_\Gamma(0)$ in terms of the first two moments of the environment-coupled sojourn time distribution, $\psi_\Gamma(t)$, as follows: $\hat{\kappa}_\Gamma(0) = \langle t \rangle_\Gamma^{-1} (\equiv k_\Gamma)$ and $\hat{\kappa}'_\Gamma(0) = 2^{-1}\left[\left(\langle t^2 \rangle_\Gamma - \langle t \rangle_\Gamma^2\right)/\langle t \rangle_\Gamma^2 - 1\right] (\equiv 2^{-1} R_{t,\Gamma})$. Here, $k_\Gamma$ and $R_{t,\Gamma}$ represent the inverse mean and the randomness of the sojourn time when the environment is at state $\Gamma$. In equation (3a), $\bar{D}$ and $\Delta_c$ are given by $\bar{D} = (\varepsilon^2/2d)\langle k_\Gamma \rangle$ and $\Delta_c = \varepsilon^2 \langle R_{t,\Gamma} \rangle/2$, respectively. This result tells us that a finite $\Delta_c$ results from $\langle R_{t,\Gamma} \rangle$, which does not vanish whenever $\psi_\Gamma(t)$ is a non-exponential function of time. When the Poisson jump process with $\psi_\Gamma(t) = k\exp(-kt)$, we obtain $\langle t^2 \rangle_\Gamma = 2\langle t \rangle_\Gamma^2 = 2k^{-2}$, and the mean square displacement (MSD) is directly proportional to time, that is, $\Delta_c = \langle R_{t,\Gamma} \rangle = 0$ in equation (3a). This is also the case for the stochastic diffusivity model in ref. 2. For any stochastic diffusivity model, the jump process is given by a generalized Poisson process with $\psi_\Gamma(t) = k_\Gamma \exp(-k_\Gamma t)$, for which the randomness, $R_{t,\Gamma}$, of the sojourn time vanishes regardless of environmental state $\Gamma$.

**Supplementary Note 3 | Derivation of equation (5).**

Disordered fluids exhibit a universal behaviour in their transport dynamics, that is, ballistic motion at short times, diffusive motion at long times, and sub-diffusive motion at intermediate times. This universality leads us to assume that the anomalous time-dependence of the MSD observed for a variety of disordered fluids has a universal functional form. Given that this assumption is valid, the analytic expression for the MSD of a polymer fluid should also provide a quantitative explanation of the MSD of other disordered fluids. Figure 1a and Supplementary Fig. 1 demonstrate that this is indeed the case for supercooled water. In liquid water, water molecules interact with each other through hydrogen bond networks. As temperature decreases, the strength of individual hydrogen bonds grows stronger and the lifetime of hydrogen bonds increases. At supercooled temperatures, a hydrogen bond network composed of water molecules can be viewed as a polymer with slowly fluctuating bond strength and connectivity. This is why the transport dynamics of a water molecule in a hydrogen bond network is qualitatively similar to viscoelastic motion of a bead in a polymer network. Both a bead in a polymer network and a water in a hydrogen bond network have transport dynamics that are composed of an unbound mode and multiple bounds mode. Equation (5) also provides a quantitative explanation of the MSD of dense, hard-disc fluids as demonstrated in Supplementary Fig. 1. We propose equation (5) as a universal, analytic expression for the MSD of disordered fluids.

Here, we provide the derivation of equation (5) for the MSD of a bead in a polymer with an arbitrary topological connectivity. Let us consider a polymer consisting of $n+1$ beads with potential energy given by

$$U/k_B T = \frac{3}{2b^2} \sum_{\alpha=x,y,z} \mathbf{X}_\alpha^T \cdot \mathbf{A} \cdot \mathbf{X}_\alpha , \qquad (\text{N3-1})$$

where $b$ is the root-mean-squared bond length between two adjacent beads, and $\mathbf{X}_\alpha$ represents the $(n+1)$-dimensional column vector defined by

$$\mathbf{X}_x = \begin{pmatrix} x_0 \\ x_1 \\ \vdots \\ x_n \end{pmatrix}, \quad \mathbf{X}_y = \begin{pmatrix} y_0 \\ y_1 \\ \vdots \\ y_n \end{pmatrix}, \quad \mathbf{X}_z = \begin{pmatrix} z_0 \\ z_1 \\ \vdots \\ z_n \end{pmatrix} \tag{N3-2}$$

with $(x_i, y_i, z_i)$ being a position of the $i$th bead. In equation (N3-1), the superscript $T$ denotes the transpose. $3k_B T \mathbf{A}/b^2$ is the Hessian matrix of the polymer. Because $\mathbf{A}$ is a $(n+1)\times(n+1)$ symmetric matrix, it can be transformed into a diagonal matrix, $\mathbf{\Lambda}$, by

$$\mathbf{A} = \mathbf{Q} \cdot \mathbf{\Lambda} \cdot \mathbf{Q}^T, \tag{N3-3}$$

where $\mathbf{Q}$ is a diagonal matrix satisfying $\mathbf{Q}^T \cdot \mathbf{Q} = \mathbf{Q} \cdot \mathbf{Q}^T = \mathbf{I}$ with $\mathbf{I}$ being the $(n+1)\times(n+1)$ identity matrix. Diagonal elements of $\mathbf{\Lambda}$ are the eigenvalues of $\mathbf{A}$, that is, $(\mathbf{\Lambda})_{ij} = \lambda_i \delta_{ij}$. $\mathbf{A}$ has a single zero eigenvalue and $n$ positive eigenvalues, i.e. $\lambda_0 (=0) < \lambda_1 \leq \lambda_2 \leq \cdots \leq \lambda_n$.

When a polymer is embedded in a three-dimensional isotropic medium, the Langevin equation of $\mathbf{X}_\alpha$ is given by

$$M\ddot{\mathbf{X}}_\alpha = -\zeta \dot{\mathbf{X}}_\alpha - \frac{3k_B T}{b^2}\mathbf{A}\cdot\mathbf{X}_\alpha + \mathbf{f}_\alpha(t), \qquad (\alpha \in \{x,y,z\}) \tag{N3-4}$$

where $M$ and $\zeta$ denote the mass and friction coefficient of a single bead, respectively. The upper dot and double dot on top of $\mathbf{X}_\alpha$ in equation (N3-4) respectively denote the first- and second-order time derivatives. On the right-hand side of equation (N3-4), the second term is an $(n+1)$-dimensional column vector, the $i$th element of which is the systematic force exerted by the potential energy, equation (N3-1), on the *i*th bead along $\alpha$ axis. The third term, $\mathbf{f}_\alpha(t)$, represents a random fluctuating force. More specifically, $\mathbf{f}_\alpha(t)$ is a $(n+1)$-dimensional column vector defined by

$$\mathbf{f}_x = \begin{pmatrix} f_{x0} \\ f_{x1} \\ \vdots \\ f_{xn} \end{pmatrix}, \quad \mathbf{f}_y = \begin{pmatrix} f_{y0} \\ f_{y1} \\ \vdots \\ f_{yn} \end{pmatrix}, \quad \mathbf{f}_z = \begin{pmatrix} f_{z0} \\ f_{z1} \\ \vdots \\ f_{zn} \end{pmatrix}, \quad \text{(N3-5)}$$

where $f_{\alpha i}$ is a random force satisfying the zero mean and the white noise correlation:

$$\langle f_{\alpha i}(t) \rangle = 0, \quad \text{(N3-6a)}$$

$$\langle f_{\alpha i}(t) f_{\beta j}(t') \rangle = 2k_B T \zeta \delta_{\alpha\beta} \delta_{ij} \delta(t-t'). \quad \alpha, \beta \in \{x, y, z\} \quad \text{(N3-6b)}$$

We can rewrite equation (N3-4) as Langevin equations for $n+1$ independent normal coordinates, defined by $\mathbf{q}_\alpha = \mathbf{Q}^T \cdot \mathbf{X}_\alpha$. This normal mode representation of equation (N3-4) can be obtained by multiplying both sides of equation (N3-4) by $\mathbf{Q}^T$ from the left:

$$M\ddot{\mathbf{q}}_\alpha = -\zeta \dot{\mathbf{q}}_\alpha - \frac{3k_B T}{b^2} \mathbf{\Lambda} \cdot \mathbf{q}_\alpha + \mathbf{\eta}_\alpha, \quad \text{(N3-7)}$$

where $\mathbf{\eta}_\alpha$ is the $(n+1)$-dimensional column vector, the *i*th element of which is the $\alpha$ component of the random force exerted on the *i*th normal coordinate, defined by $\mathbf{\eta}_\alpha = \mathbf{Q}^T \cdot \mathbf{f}_\alpha$. Here, the elements of $\mathbf{\eta}_\alpha$ also satisfy equation (N3-6), that is,

$$\langle \eta_{\alpha i}(t) \rangle = \sum_{k=0}^{n} (\mathbf{Q}^T)_{ik} \langle f_{\alpha k}(t) \rangle = 0, \quad \text{(N3-8a)}$$

$$\langle \eta_{\alpha i}(t)\eta_{\beta j}(t')\rangle = \sum_{k=0}^{n}\sum_{l=0}^{n}(\mathbf{Q}^T)_{ik}(\mathbf{Q}^T)_{il}\langle f_{\alpha k}(t)f_{\beta l}(t)\rangle$$

$$= 2k_B T\zeta\delta_{\alpha\beta}\delta(t-t')\sum_{k=0}^{n}\sum_{l=0}^{n}(\mathbf{Q}^T)_{ik}(\mathbf{Q}^T)_{jl}\delta_{kl} \qquad \text{(N3-8b)}$$

$$= 2k_B T\zeta\delta_{\alpha\beta}\delta(t-t')\sum_{k=0}^{n}(\mathbf{Q}^T)_{ik}(\mathbf{Q})_{kj}$$

$$= 2k_B T\zeta\delta_{\alpha\beta}\delta_{ij}\delta(t-t').$$

Therefore, the time evolution of the $n+1$ normal modes are independent of each other. For the zeroth mode, equation (N3-7) gives

$$\ddot{q}_{\alpha 0} = -\gamma \dot{q}_{\alpha 0} + \eta_{\alpha 0}/M \qquad \text{(N3-9)}$$

with $\gamma$ being the velocity relaxation rate defined by $\gamma = \zeta/M$. Otherwise, i.e. $i \geq 1$, we have

$$\ddot{q}_{\alpha i} = -\gamma \dot{q}_{\alpha i} - \omega_{0,i}^2 q_{\alpha i} + \eta_{\alpha i}/M . \qquad \text{(N3-10)}$$

with $\omega_{0,i}$ being the natural frequency of the $i$th bound mode defined by $M\omega_{0,i}^2 = 3k_B T\lambda_i/b^2$.

Using the relation, $\mathbf{q}_\alpha = \mathbf{Q}^T \cdot \mathbf{X}_\alpha$ or $\mathbf{X}_\alpha = \mathbf{Q} \cdot \mathbf{q}_\alpha$, the MSD of the $i$th bead can be written as

$$\langle[\mathbf{r}_i(t)-\mathbf{r}_i(0)]^2\rangle = \sum_{\alpha=x,y,z}\sum_{k=0}^{n}\sum_{l=0}^{n}(\mathbf{Q})_{ik}(\mathbf{Q})_{il}\langle[q_{\alpha k}(t)-q_{\alpha k}(0)][q_{\alpha l}(t)-q_{\alpha l}(0)]\rangle$$
$$= \sum_{\alpha=x,y,z}\sum_{k=0}^{n}(\mathbf{Q})_{ik}^2\langle[q_{\alpha k}(t)-q_{\alpha k}(0)]^2\rangle. \qquad \text{(N3-11)}$$

To calculate equation (N3-11), we need the MSD of the $k$th mode, which can be obtained with equations (N3-9) and (N3-10). The ordinary Langevin equation, (N3-10), of a free particle can be easily solved with respect to $\dot{q}_{\alpha 0}$:

$$\dot{q}_{\alpha 0}(t) = e^{-t\gamma}\dot{q}_{\alpha 0}(0) + \int_0^t dt' e^{-(t-t')\gamma}\eta_{\alpha 0}(t')/M . \qquad \text{(N3-12)}$$

Using equation (N3-12) and noting that $q_{\alpha 0}(t) - q_{\alpha 0}(0) = \int_0^t dt' \dot{q}_{\alpha 0}(t')$, the MSD of the zeroth mode can then be calculated as [3]

$$\langle [q_{\alpha 0}(t) - q_{\alpha 0}(0)]^2 \rangle = \int_0^t dt_1 \int_0^t dt_2 \langle \dot{q}_{\alpha 0}(t_1) \dot{q}_{\alpha 0}(t_2) \rangle$$
$$= 2 \frac{k_B T}{M \gamma^2} (\gamma t - 1 + e^{-\gamma t}), \qquad \text{(N3-13)}$$

where we have used $\langle \dot{q}_{\alpha 0}^2 \rangle = k_B T / M$ for the initial Maxwell-Boltzmann distribution and the orthogonality condition, $\langle \eta_{\alpha 0}(t) \dot{q}_{\alpha 0}(0) \rangle = 0$. After some rearrangement, the MSD of the $k$th mode ($k \geq 1$) can be rewritten as

$$\langle [q_{\alpha k}(t) - q_{\alpha k}(0)]^2 \rangle = 2 \langle q_{\alpha k}^2 \rangle \left[ 1 - \phi_{q_{\alpha k}}(t) \right] \qquad \text{(N3-14)}$$

with the initial equilibrium condition. In equation (N3-14), $\langle q_{\alpha k}^2 \rangle$ denotes the mean square of $q_{\alpha k}$ at equilibrium, which is given by $\langle q_{\alpha k}^2 \rangle = k_B T / M \omega_{0,k}^2$. $\phi_{q_{\alpha k}}(t)$ $\left( = \langle q_{\alpha k}(t) q_{\alpha k}(0) \rangle / \langle q_{\alpha k}^2 \rangle \right)$ is the normalized time correlation function of $q_{\alpha k}$. $\phi_{q_{\alpha k}}(t)$ satisfies the following equation:

$$\ddot{\phi}_{q_{\alpha k}} = -\gamma \dot{\phi}_{q_{\alpha k}} - \omega_{0,i}^2 \phi_{q_{\alpha k}}, \qquad \text{(N3-15)}$$

which is obtained from equation (N3-10) with the orthogonality condition, $\langle \eta_{\alpha k}(t) q_{\alpha k}(0) \rangle = 0$. The ordinary second-order differential equation, equation (N3-15), can be easily solved with two initial conditions, $\phi_{q_{\alpha k}}(0) = 1$ and $\dot{\phi}_{q_{\alpha k}}(0) = 0$. The condition $\dot{\phi}_{q_{\alpha k}}(0) = 0$ is needed for the MSD of the $i$th bead to be quadratic in time at short times. The explicit expression of $\phi_{q_{\alpha k}}(t)$ is given by

$$\phi_{q_{\alpha k}}(t) = e^{-\gamma t/2} \left[ \cosh(\omega_k t) + \frac{\gamma}{2\omega_k} \sinh(\omega_k t) \right] \qquad \text{(N3-16)}$$

with $\omega_k = \sqrt{(\gamma/2)^2 - \omega_{0,k}^2}$. With equations (N3-13), (N3-14), and (N3-16) at hand, equation (N3-11) can be rewritten as

$$\langle [\mathbf{r}_i(t) - \mathbf{r}_i(0)]^2 \rangle = 6 \frac{k_B T}{M \gamma^2} c_0 (\gamma t - 1 + e^{-\gamma t})$$
$$+ 6 \frac{k_B T}{M} \sum_{k=1}^{n} \frac{c_k}{\omega_{0,k}^2} \left[ 1 - e^{-\gamma t/2} \left( \cosh \omega_k t + \frac{\gamma}{2\omega_k} \sinh \omega_k t \right) \right], \tag{N3-17}$$

where $c_k \left[ \equiv (\mathbf{Q})_{ik}^2 \right]$ denotes the normalized weight coefficient for the $k$th mode. The diffusive version of equation (N3-17) can be found in ref. [4].

Let $\gamma_k$ denote the relaxation rate for the $k$th mode. Equation (N3-17) can then be regarded as a special case in which $\gamma_0 = \gamma$ and $\gamma_k = \gamma/2$. In general, $\gamma_k$ can differ from mode to mode [5]. With the $k$-dependent mode relaxation rates, $\{\gamma_k\}$, equation (N3-17) can be rewritten as

$$\langle [\mathbf{r}_i(t) - \mathbf{r}_i(0)]^2 \rangle = 6 \frac{k_B T}{M \gamma_0^2} c_0 (\gamma_0 t - 1 + e^{-\gamma_0 t})$$
$$+ 6 \frac{k_B T}{M} \sum_{k=1}^{n} \frac{c_k}{\omega_{0,k}^2} \left[ 1 - e^{-\gamma_k t} \left( \cosh \omega_k t + \frac{\gamma_k}{\omega_k} \sinh \omega_k t \right) \right] \tag{N3-18}$$

with $\omega_k = \sqrt{\gamma_k^2 - \omega_{0,k}^2}$. Equation (5) given in the main text is the $d$-dimensional version of equation (N3-18). Equation (5) or (N3-18) can be generalized to include weakly damped modes characterized by $\gamma_k < \omega_{0,k}$, whose contributing terms show an oscillatory behavior over time. In applying equation (5) to supercooled TIP4P/2005 water and dense hard-disc systems, we assume only overdamped modes ($\gamma_k > \omega_{0,k}$) in equation (5).

**Supplementary Note 4 | Short-time limiting behaviour of the non-Gaussian parameter.**

In this note, we provide a brief discussion of the short-time limiting value of the non-Gaussian parameter (NGP), which complements the long-time limiting value of the NGP on the basis of equation (2b). Without loss of generality, the short-time limiting behaviour of the MSD can be represented by $\Delta_2(t) \cong ct^\beta$ ($\beta > 0$), whose Laplace transform is given by $\hat{\Delta}_2(s) \cong cs^{-1-\beta}\Gamma(1+\beta)$ with $\Gamma(z)$ denoting the Gamma function defined by $\Gamma(z) = \int_0^\infty dt\, t^{z-1} e^{-t}$. Substituting $\hat{\Delta}_2(s) \cong cs^{-1-\beta}\Gamma(1+\beta)$ into equation (2b) and taking the large-$s$ limit of the resulting equation, we have

$$\hat{\Delta}_4(s) \cong \left(1 + \frac{2}{d}\right) 2c^2 s^{-1-2\beta} \Gamma(1+\beta)^2 [1 + C_{\mathcal{D}}(0)], \tag{N4-1}$$

where we have used the Tauberian theorem, i.e. $\lim_{s \to \infty} s\hat{C}_{\mathcal{D}}(s) = \lim_{t \to 0} C_{\mathcal{D}}(t) = C_{\mathcal{D}}(0)$. The inverse Laplace transform of equation (N2-1) is then given by

$$\Delta_4(t) \cong \left(1 + \frac{2}{d}\right) \frac{2c^2 \Gamma(1+\beta)^2}{\Gamma(1+2\beta)} [1 + C_{\mathcal{D}}(0)] t^{2\beta}. \tag{N4-2}$$

Substituting equation (N4-2) and $\Delta_2(t) \cong ct^\beta$ into the definition of the NGP, equation (40), the short-time limiting value of the NGP can be obtained as

$$\alpha_2(0) = \frac{2\Gamma(1+\beta)^2}{\Gamma(1+2\beta)} [1 + C_{\mathcal{D}}(0)] - 1, \tag{N4-3}$$

which reduces to

$$\alpha_2(0) = [C_{\mathcal{D}}(0) - 2]/3 \tag{N4-4}$$

for $\beta = 2$. Given that the initial velocity follows the Maxwell-Boltzmann distribution, the values of $\beta$ and $C_\mathcal{D}(0)$ are respectively given by $\beta = 2$ and $C_\mathcal{D}(0) = 2$, which is discussed in the main text. In this case, the initial value of the NGP, equation (N2-4), is zero, which is consistent with the simulation results shown in Fig. 3b and Supplementary Fig. 1b.

For other transport models, the initial value of the NGP is nonzero. For the stochastic diffusivity (SD) model, the value of $\beta$ is unity and $C_\mathcal{D}(t=0)$ is replaced by $\eta_D^2 \phi_D(t=0) = \eta_D^2$, yielding the following result: $\alpha_2(0) = \eta_D^2$, which is consistent with equation (45) in Methods. For the environment-invariant model characterized by $C_\mathcal{D}(t) = 0$, the initial value of the NGP is given by $\alpha_2(0) = -2/3$.

When the MSD is given by equation (5), we can discuss the next leading term of the NGP in addition to the initial value of the NGP. At short times, equation (5) can be expanded as

$$\Delta_2(t)/2d\langle v_\alpha^2 \rangle = \frac{1}{2}t^2 - \frac{1}{6}\gamma t^3 + \frac{1}{24}\mu t^4 + \mathcal{O}(t^5) \tag{N4-5}$$

with $\langle v_\alpha^2 \rangle = k_B T/M$. In equation (N4-5), $\gamma$ and $\mu$ are given by $\gamma = c_0 \gamma_0 + 2\sum_{i=1}^{n} c_i \gamma_i$ and $\mu = c_0 \gamma_0^2 + \sum_{i=1}^{n} c_i (3\gamma_i^2 + \omega_i^2)$, respectively. In addition, the short-time behavior of $C_\mathcal{D}(t)$ is described by equation (32) as shown in Fig. 2a and Supplementary Fig. 1a, explicitly, $C_\mathcal{D}(t) \cong 2\phi_{v_\alpha}(t)^2$ at short times. The analytical expression of the TCF, $\phi_{v_\alpha}(t)$, of $v_{\alpha \in \{x,y,z\}}$ can be obtained by differentiating equation (5) two times with respect to time. The resulting equation is given by

$$\phi_{v_\alpha}(t) = c_0 e^{-\gamma_0 t} + \sum_{i=1}^{n} c_i e^{-\gamma_i t}\left(\cosh \omega_i t - \frac{\gamma_i}{\omega_i}\sinh \omega_i t\right) \tag{N4-6}$$

whose initial decay rate is given by $\gamma$. We can then obtain the short-time expansion of $C_D(t)$, which is given by

$$C_D(t)/2 = 1 - 2\gamma t + (\gamma^2 + \mu)t^2 + \mathcal{O}(t^3) \tag{N4-7}$$

Substituting the Laplace transforms of equations (N4-6) and (N4-7) into equation (2b), the inverse Laplace transformation of the resulting expression of $\hat{\Delta}_4(s)$ gives the short-time expansion of $\Delta_4(t)$:

$$\Delta_4(t) = (2d\langle v_\alpha^2 \rangle)^2 \left(1 + \frac{2}{d}\right)\left[\frac{1}{4}t^4 - \frac{1}{6}\gamma t^5 + \left(\frac{\gamma^2}{24} + \frac{\mu}{36}\right)t^6 + \mathcal{O}(t^7)\right] \tag{N4-8}$$

Substituting equations (N4-6) and (N4-8) into the definition of the NGP, $\alpha_2(t)\left[=\Delta_4(t)/\left[(1+2/d)\Delta_2(t)^2\right]-1\right]$, we finally have

$$\alpha_2(t) = \frac{1}{18}(\gamma^2 - \mu)t^2 + \mathcal{O}(t^3) \tag{N4-9}$$

To understand which sign $\gamma^2 - \mu$ has on the R.H.S. of equation (N4-9), let's look into equation (N4-5). When $\mu$ is replaced by $\gamma^2$ on the R.H.S. of equation (N4-5), in other words, $\mu = \gamma^2$, equation (N4-5) is just a short-time expansion of $(\gamma t - 1 + e^{-\gamma t})/\gamma^2$. Note here that the MSD given by $\Delta_2(t) = 2d\langle v_\alpha^2 \rangle(\gamma t - 1 + e^{-\gamma t})/\gamma^2$ corresponds to the upper limit of equation (5) with the nonzero bound mode contribution as shown in Fig. 2a or Supplementary Fig. 1a. Therefore, one can conclude that $\mu$ is smaller than $\gamma^2$ when the MSD shows a subdiffusive behaviour around the caging time, i.e., $\gamma^2 - \mu > 0$, which is consistent with the short-time behaviour of the NGP as shown in Fig. 2b or Supplementary Figs. 1b, 2b, and 3.

**Supplementary Note 5 | Relationship between displacement statistics and jump number statistics.**

In this note, we present the mapping of jump number on lattice and displacement statistics. Let $m_i$ denote the on-lattice coordinate of a random walker along the $i$th axis in $d$-dimensional space. Given that a random walker is initially located at the coordinate origin, the second and fourth moments of the displacement, $r(t)$, of the random walker at time $t$ are then given by

$$\langle r^2(t) \rangle \equiv \Delta_2(t) = \varepsilon^2 \sum_{i=1}^{d} \langle m_i^2(t) \rangle \tag{N5-1a}$$

$$\langle r^4(t) \rangle \equiv \Delta_4(t) = \varepsilon^4 \sum_{i=1}^{d} \sum_{j=1}^{d} \langle m_i^2(t) m_j^2(t) \rangle \tag{N5-1b}$$

with $\varepsilon$ denoting a lattice constant. Here, $m_i$ is equal to the difference between the number, $n_{i,+}$, of jumps made in the positive direction and the number, $n_{i,-}$, of jumps made in the negative direction along the $i$th axis, i.e. $m_i = n_{i,+} - n_{i,-}$. Equations (N5-1a) and (N5-1b) can be rewritten as

$$\Delta_2(t) = \varepsilon^2 \sum_{N=0}^{\infty} \left[ \sum_{i=1}^{d} \langle (n_{i,+} - n_{i,-})^2 \rangle_N \right] P_N(t) \tag{N5-2a}$$

$$\begin{aligned}\Delta_4(t) &= \varepsilon^4 \sum_{N=0}^{\infty} \left[ \sum_{i=1}^{d} \sum_{j=1}^{d} \langle (n_{i,+} - n_{i,-})^2 (n_{j,+} - n_{j,-})^2 \rangle_N \right] P_N(t) \\ &= \varepsilon^4 \sum_{N=0}^{\infty} \left[ \sum_{i=1}^{d} \langle (n_{i,+} - n_{i,-})^4 \rangle_N + 2 \sum_{j>i} \langle (n_{i,+} - n_{i,-})^2 (n_{j,+} - n_{j,-})^2 \rangle_N \right] P_N(t) \end{aligned} \tag{N5-2b}$$

where $\langle \cdots \rangle_N$ and $P_N(t)$ denote, respectively, the average taken under the condition that the total number of jumps made by a random worker is $N$ and the probability that the total number of jump made by a random walker is $N$ at time $t$.

The conditional averages $\langle f(\mathbf{n})\rangle_N$ can be calculated by $\langle f(\mathbf{n})\rangle_n = \sum_{\mathbf{n}}^* f(\mathbf{n}) p(\mathbf{n}|N)$, where $\sum_{\mathbf{n}}^*(\cdots)$ and $p(\mathbf{n}|N)$ represent, respectively, the sum over all possible jump number vectors $\mathbf{n}[=(n_{1,+}, n_{1,-}, \cdots, n_{d,+}, n_{d,-})]$ satisfying $\sum_{i=1}^d (n_{i,+} + n_{i,-}) = N$, and the multinomial distribution given by

$$p(\mathbf{n}|N) = \frac{N!}{\prod_{i=1}^d n_{i,+}! n_{i,-}!} \prod_{i=1}^d p_{i,+}^{n_{i,+}} p_{i,-}^{n_{i,-}} \tag{N5-3}$$

where $p_{i,+(-)}$ is the probability of a jump in the positive (negative) direction along the $i$th axis. $\{p_{i,\pm}\}$ satisfies the normalization condition, $\sum_{i=1}^d (p_{i,+} + p_{i,-}) = 1$. Henceforth, we consider isotropic, unbiased random walks, for which the value of $p_{i,\pm}$ is given by $p_{i,\pm} = p = 1/2d$ for any $i$.

In order to calculate the conditional moments, $\langle (n_{i,+} - n_{i,-})^2 \rangle_N$, $\langle (n_{i,+} - n_{i,-})^4 \rangle_N$, and $\langle (n_{i,+} - n_{i,-})^2 (n_{j,+} - n_{j,-})^2 \rangle_N$, in equations (N5-2a) and (N5-2b), it is convenient to make use of the moment generating function, $f(\boldsymbol{\lambda}|N)$ $[\boldsymbol{\lambda} = (\lambda_{1,+}, \lambda_{1,-}, \cdots, \lambda_{d,+}, \lambda_{d,-})]$, of equation (N5-3). The moment generating function is defined by

$$f(\boldsymbol{\lambda}|N) = \sum_{\mathbf{n}}^* \lambda_{1,+}^{n_{1,+}} \lambda_{1,-}^{n_{1,-}} \lambda_{2,+}^{n_{2,+}} \lambda_{2,-}^{n_{2,-}} \cdots \lambda_{d,+}^{n_{d,+}} \lambda_{d,-}^{n_{d,-}} p(\mathbf{n}|N) \tag{N5-4}$$

where the asterisk next to the summation signifies the constraint, $\sum_{i=1}^d (n_{i,+} + n_{i,-}) = N$, and it is well-known that $f(\boldsymbol{\lambda}|N)$ is given by

$$f(\boldsymbol{\lambda}|N) = \left( p \sum_{i=1}^d \lambda_{i,+} \right)^N \tag{N5-5}$$

It is also well-known that the derivatives of $f(\boldsymbol{\lambda}|N)$ with respect to $\lambda_{i,\pm}$'s are directly related to the conditional moments:

$$\left.\frac{\partial^\alpha f}{\partial \lambda_{i,\pm}^\alpha}\right|_{\boldsymbol{\lambda}=\mathbf{1}} = \langle n_{i,\pm}(n_{i,\pm}-1)\cdots(n_{i,\pm}-\alpha+1)\rangle_N \tag{N5-6a}$$

$$\left.\frac{\partial^{\alpha+\beta} f}{\partial \lambda_{i,\pm}^\alpha \partial \lambda_{j,\pm}^\beta}\right|_{\boldsymbol{\lambda}=\mathbf{1}} = \langle n_{i,\pm}(n_{i,\pm}-1)\cdots(n_{i,\pm}-\alpha+1)n_{j,\pm}(n_{j,\pm}-1)\cdots(n_{j,\pm}-\beta+1)\rangle_N \tag{N5-6b}$$

$$\left.\frac{\partial^{2+\alpha} f}{\partial \lambda_{i,+} \partial \lambda_{i,-} \partial \lambda_{j,\pm}^\alpha}\right|_{\boldsymbol{\lambda}=\mathbf{1}} = \langle n_{i,+}n_{i,-}n_{j,\pm}(n_{j,\pm}-1)\cdots(n_{j,\pm}-\alpha+1)\rangle_N \tag{N5-6c}$$

$$\left.\frac{\partial^4 f}{\partial \lambda_{i,+} \partial \lambda_{i,-} \partial \lambda_{j,+} \partial \lambda_{j,-}}\right|_{\boldsymbol{\lambda}=\mathbf{1}} = \langle n_{i,+}n_{i,-}n_{j,+}n_{j,-}\rangle_N \tag{N5-6d}$$

where $\alpha$ and $\beta$ are positive integers. On the left-hand side of equation (N5-6), $\mathbf{1}$ designates the $2d$-dimensional row vector, every element of which is unity. Substituting equation (N5-5) into equation (N5-6), we obtain

$$\langle n_{i,\pm}(n_{i,\pm}-1)\cdots(n_{i,\pm}-\alpha+1)\rangle_N = N(N-1)\cdots(N-\alpha+1)p^\alpha \tag{N5-7a}$$

$$\langle n_{i,\pm}(n_{i,\pm}-1)\cdots(n_{i,\pm}-\alpha+1)n_{j,\pm}(n_{j,\pm}-1)\cdots(n_{j,\pm}-\beta+1)\rangle_N$$
$$= N(N-1)\cdots(N-\alpha-\beta+1)p^{\alpha+\beta} \tag{N5-7b}$$

$$\langle n_{i,+}n_{i,-}n_{j,\pm}(n_{j,\pm}-1)\cdots(n_{j,\pm}-\alpha+1)\rangle_N = N(N-1)\cdots(N-\alpha-1)p^{2+\alpha} \tag{N5-7c}$$

$$\langle n_{i,+}n_{i,-}n_{j,+}n_{j,-}\rangle_N = N(N-1)(N-2)(N-3)p^4 \tag{N5-7d}$$

From equation (N5-7), one can obtain $\langle(n_{i,+}-n_{i,-})^2\rangle_N$, $\langle(n_{i,+}-n_{i,-})^4\rangle_N$, and $\langle(n_{i,+}-n_{i,-})^2(n_{j,+}-n_{j,-})^2\rangle_N$:

$$\langle(n_{i,+}-n_{i,-})^2\rangle_N = 2\langle n_{i,\pm}^2\rangle - 2\langle n_{i,+}n_{i,-}\rangle = n/d \tag{N5-8a}$$

$$\begin{aligned}\langle(n_{i,+}-n_{i,-})^4\rangle_N &= 2\langle n_{i,\pm}^4\rangle - 8\langle n_{i,\pm}^3 n_{i,\mp}\rangle + 6\langle n_{i,+}^2 n_{i,-}^2\rangle \\ &= 3n(n-1)/d^2 + n/d\end{aligned} \tag{N5-8b}$$

$$\begin{aligned}\langle(n_{i,+}-n_{i,-})^2(n_{j,+}-n_{j,-})^2\rangle_N &= 2\langle n_{i,\pm}^2 n_{j,\pm}^2\rangle + 2\langle n_{i,\pm}^2 n_{j,\mp}^2\rangle \\ &\quad - 2\langle n_{i,+}n_{i,-}n_{j,\pm}^2\rangle - 2\langle n_{i,\pm}^2 n_{j,+}n_{j,-}\rangle + 4\langle n_{i,+}n_{i,-}n_{j,+}n_{j,-}\rangle \\ &= n(n-1)/d^2\end{aligned} \tag{N5-8c}$$

Substituting equation (N5-8) into equation (N5-2), we obtain the analytic expressions for the mean square displacement (MSD), the fourth moment, and the non-Gaussian parameter (NGP) as

$$\Delta_2(t) = \varepsilon^2 \langle N(t)\rangle \tag{N5-9}$$

$$\Delta_4(t) = \varepsilon^4 \left[\left(1+\frac{2}{d}\right)\langle N(N-1)(t)\rangle + \langle N(t)\rangle\right] \tag{N5-10}$$

$$\alpha_2(t) = \frac{\Delta_4(t)}{3\Delta_2(t)^2} - 1 = \frac{1}{\langle N(t)\rangle}\left(Q_N(t) + \frac{d}{d+2}\right) \tag{N5-11}$$

where $\langle N^k(t)\rangle$ and $Q_N(t)$ denote the $k$th moment of jump number $N$, defined by $\langle N^k(t)\rangle = \sum_{N=0}^{\infty} N^k P_N(t)$, and the Mandel's Q parameter[6] of the jump number defined by $Q_N(t) = (\langle N^2(t)\rangle - \langle N(t)\rangle^2)/\langle N(t)\rangle - 1$.

**Supplementary Note 6 | Description of the enzyme reaction-coupled transport model in Supplementary Fig. 2.**

In this note, we present the analytical expressions of the MSD and NGP along with the simulation method used in Supplementary Fig. 2. For the enzyme reaction-coupled transport model shown in Supplementary Fig. 2a, where a single enzyme turnover event leads to a single random step of the motor enzyme along a one-dimensional track, we use equations (N5-9) and (N5-11) to calculate the MSD and NGP of motor enzyme displacement. Note that equations (N5-9) and (N5-11) relate the MSD and NGP to the first two moments of the jump number distribution, $\langle N(t) \rangle [\equiv \mu_1(t)]$ and $\langle N^2(t) \rangle$ or $\langle N(N-1)(t) \rangle [\equiv \mu_2(t)]$. The Laplace domain expressions of $\mu_1(t)$ and $\mu_2(t)$ can be obtained by using equations (23) and (26) in Method as follows [7,8]:

$$\hat{\mu}_1(s) = \frac{1}{s^2} \int d\Gamma \, \hat{\kappa}_\Gamma(s) p_{st}(\Gamma) \tag{N6-1a}$$

$$\hat{\mu}_2(s) = \frac{2}{s^2} \int d\Gamma \int d\Gamma_0 \, \hat{\kappa}_\Gamma(s) G(\Gamma, s | \Gamma_0) \hat{\kappa}_{\Gamma_0}(s) p_{st}(\Gamma_0) \tag{N6-1b}$$

Here, the enzyme state-dependent rate kernel, $\hat{\kappa}_\Gamma(s)$, is related to the enzyme state-dependent sojourn time distribution, $\hat{\psi}_\Gamma(s)$, by

$$\hat{\kappa}_\Gamma(s) = \frac{s \hat{\psi}_\Gamma(s)}{1 - \hat{\psi}_\Gamma(s)} \tag{N6-2}$$

For the model considered in Supplementary Fig. 2, the enzyme state-dependent sojourn time distribution, $\hat{\psi}_\Gamma(s)$, in the Laplace domain can be obtained as [8]

$$\hat{\psi}_\Gamma(s) = \frac{p_2(\Gamma) \hat{\varphi}_1(s) \hat{\varphi}_{ES}(s)}{1 - p_{-1}(\Gamma) \hat{\varphi}_1(s) \hat{\varphi}_{ES}(s)} \tag{N6-3}$$

where $\hat{\varphi}_1(s)$ or $\hat{\varphi}_{ES}(s)$ denotes the Laplace transform of the distribution, $\varphi_1(t)$, of time elapsed to complete an enzyme-substrate association reaction or the Laplace transform of the lifetime distribution, $\varphi_{ES}(t)$, of the enzyme-substrate complex. In equation (N6-3), $p_2(\Gamma)$ or $p_{-1}(\Gamma)[=1-p_2(\Gamma)]$ denotes the enzyme state-dependent probability for an ES complex to undergo a catalytic reaction $(ES \rightarrow E+P)$ or complex dissociation reaction $(E+S \leftarrow ES)$. Substituting equation (N6-3) into equation (N6-2), we obtain the expression of the rate kernel as follows:

$$\hat{\kappa}_\Gamma(s) = p_2(\Gamma)\frac{s\hat{\varphi}_1(s)\hat{\varphi}_{ES}(s)}{1-\hat{\varphi}_1(s)\hat{\varphi}_{ES}(s)} \tag{N6-4}$$

Further substituting equation (N6-4) into equations (N6-1a) and (N6-1b), we have

$$\hat{\mu}_1(s) = \frac{\langle p_2 \rangle}{s}\frac{\hat{\varphi}_1(s)\hat{\varphi}_{ES}(s)}{1-\hat{\varphi}_1(s)\hat{\varphi}_{ES}(s)} \tag{N6-5a}$$

$$\hat{\mu}_2(s) = 2\left(\frac{\hat{\varphi}_1(s)\hat{\varphi}_{ES}(s)}{1-\hat{\varphi}_1(s)\hat{\varphi}_{ES}(s)}\right)^2 \int d\Gamma \int d\Gamma_0 p_2(\Gamma)\hat{G}(\Gamma,s|\Gamma_0)p_2(\Gamma_0)p_{st}(\Gamma_0). \tag{N6-5b}$$

In equation (N6-5a), $\langle p_2 \rangle$ designates the average of $p_2(\Gamma)$ over the steady-state distribution, $p_{st}(\Gamma)$, of enzyme state $\Gamma$, i.e., $\langle p_2 \rangle = \int d\Gamma p_2(\Gamma) p_{st}(\Gamma)$. Using equation (N6-5a), we rewrite equation (N6-5b) as

$$\hat{\mu}_2(s) = 2s\hat{\mu}_1(s)^2\left[1+s\eta_{p_2}^2\hat{\phi}_{p_2}(s)\right], \tag{N6-6}$$

where $\eta_{p_2}^2$ and $\hat{\phi}_{p_2}(s)$ denote, respectively, the relative variance of $p_2$ and the Laplace transform of the time correlation function, $\phi_{p_2}(t)$, of fluctuation in $p_2$, that is,

$$\phi_{p_2}(t) = \frac{\langle \delta p_2(t) \delta p_2(0) \rangle}{\langle \delta p_2^2 \rangle},$$

$$= \frac{\int d\Gamma \int d\Gamma_0 \delta p_2(\Gamma) G(\Gamma, t | \Gamma_0) \delta p_2(\Gamma_0) p_{st}(\Gamma_0)}{\int d\Gamma \delta p_2(\Gamma)^2 p_{st}(\Gamma)}.$$
(N6-7)

In equation (N6-7), $\delta p_2(\Gamma)$ denotes $p_2(\Gamma) - \langle p_2 \rangle$.

In our model calculation shown in Supplementary Fig. 2, $\hat{\varphi}_1(s)$ and $\hat{\varphi}_{ES}(s)$ are modeled by $\hat{\varphi}_1(s) = (1 + s/k_1[S])^{-1}$ and $\hat{\varphi}_{ES}(s) = (1 + s\langle t_{ES}\rangle)^{-1}$, respectively. Here, $k_1$, $[S]$, and $\langle t_{ES} \rangle$ denote the enzyme-substrate association rate, the substrate concentration, and the mean lifetime of the ES complex, respectively. For this model, the time domain expressions of equations (N6-5a) and (N6-6) can be obtained as

$$\langle N(\tilde{t}) \rangle = \langle p_2 \rangle \frac{x}{(x+1)^2} \left[ (1+x)\tilde{t} - 1 + e^{-\tilde{t}(1+x)} \right],$$
(N6-8a)

$$\langle N(N-1)(\tilde{t}) \rangle = \langle p_2 \rangle^2 g(x,\tilde{t}) + \langle p_2 \rangle^2 \eta_{p_2}^2 \int_0^{\tilde{t}} d\tilde{t}' \dot{g}(x, \tilde{t} - \tilde{t}') \phi_{p_2}(\tilde{t}'),$$
(N6-8b)

where $\tilde{t}$ and $x$ respectively denote the dimensionless time and substrate concentration given by $\tilde{t} = t/\langle t_{ES} \rangle$ and $x = k_1 \langle t_{ES} \rangle [S]$. In equation (N7-8b), $g(x,\tilde{t})$, and $\dot{g}(x,\tilde{t})$ are defined by

$$g(x,\tilde{t}) = \frac{x^2}{(x+1)^4} \left[ (1+x)^2 \tilde{t}^2 - 2(2 + e^{-\tilde{t}(1+x)})(1+x)\tilde{t} + 6(1 - e^{-\tilde{t}(1+x)}) \right],$$
(N6-9)

and $\dot{g}(x,\tilde{t}) \equiv \partial g(x,\tilde{t})/\partial \tilde{t}$, respectively.

The analytic expressions of the MSD and NGP of the models considered in Supplementary Fig. 2a are given by equations (N5-9) and (N5-11), where $\langle N(t) \rangle$ and $Q_N(t) \left[ = \left( \langle N(N-1)(t) \rangle - \langle N(t) \rangle^2 \right) / \langle N(t) \rangle \right]$ can be calculated by equations (N6-8a) and (N7-8b). To perform an explicit calculation by using (N6-8a) and (N6-8b), we also need the expression of the time correlation function, $\phi_{p_2}(t)$, and the mean and relative variance, $\langle p_2 \rangle$ and $\eta_{p_2}^2$, of $p_2$. For the three-state model shown in the inset of Supplementary Fig. 2a, $\phi_{p_2}(t)$ is given by [8]

$$\phi_{p_2}(t) = c_+ e^{-tk_{12}\lambda_+} + c_- e^{-tk_{12}\lambda_-}, \tag{N6-10}$$

where $c_+ + c_- = 1$ and $\lambda_\pm = 1 + r \pm D$ with $r$ and $D$ denoting $r = k_{23}/k_{12}$ and $D = \sqrt{1 - r + r^2}$, respectively. In equation (N6-10), $c_+$ or $c_-$ is given by

$$c_+ = 1 - c_- = \frac{1}{6\langle \delta p_2^2 \rangle}\left[a_1(p_{2,1} - p_{2,2})^2 + a_2(p_{2,2} - p_{2,3})^2 + a_3(p_{2,3} - p_{2,1})^2\right], \tag{N6-11}$$

where $a_1 = (1+D)/(\lambda_+ D)$, $a_2 = 1 - a_1 - a_3$, and $a_3 = -r/(\lambda_+ D)$. In equation (N6-11), $p_{2,i}$ denotes $p_2(\Gamma_i)$; for the model shown in the inset of Supplementary Fig. 2a, we have $p_{2,1} = 1$, $p_{2,2} = 0.6$, and $p_{2,3} = 0.3$. For this three-state model, where the forward and backward transition rates between a pair of enzyme states are the same, the stationary probability of every enzyme state is given by 1/3; therefore, the mean, variance, and relative variance of $p_2$ are given by $\langle p_2 \rangle = \sum_{i=1}^{3} p_{2,i}/3 = 0.6\dot{3}$, $\langle \delta p_2^2 \rangle = \sum_{i=1}^{3} \delta p_{2,i}^2/3 = 0.12\dot{3}$, and $\eta_{p_2}^2 = \langle \delta p_2^2 \rangle / \langle p_2 \rangle^2 \cong 0.307$, respectively. For the non-ergodic case where $r = k_{23}/k_{12} = 0$, equation (N6-10) reduces to

$$\phi_{p_2}(t) = \frac{(p_{2,1} - p_{2,2})^2}{6\langle \delta p_2^2 \rangle} e^{-2k_{12}t} + \left[1 - \frac{(p_{2,1} - p_{2,2})^2}{6\langle \delta p_2^2 \rangle}\right], \tag{N6-12}$$

whose long-time limit value does not vanish. The result for the non-ergodic case is represented by the red line in Supplementary Fig. 2c-f.

For the model considered in Supplementary Fig. 2b, the MSD and NGP can be easily calculated by substituting equations (N6-8a) and (N6-8b) with $\eta_{p_2}^2 = 0$ into equations (N5-9) and (N5-11). In the absence of fluctuation in $p_2$, i.e. $\eta_{p_2}^2 = 0$, equations (N6-8a) and (N6-8b) conform to the result of the CTRW model, namely, $\hat{\mu}_2(s) = 2s\hat{\mu}_1(s)^2$.

Values of the NGP calculated with equation (N5-11) for the model shown in Supplementary Fig. 2a approach values of the NGP calculated with equation (2b) in the continuum limit. To show this, we rewrite equation (N5-10) as

$$\Delta_4(t) = \varepsilon^4 \left(1 + \frac{2}{d}\right)\langle N(N-1)(t)\rangle + \varepsilon^2 \Delta_2(t) \tag{N6-13}$$

using equation (N5-9). Substituting equation (N6-13) into the definition of the NGP, we obtain equation (N5-11), which generally yields different results from equation (2b). However, in the small $\varepsilon$ limit, equation (N6-13) reduces to

$$\Delta_4(t) = \varepsilon^4 \left(1 + \frac{2}{d}\right)\mu_2(t), \tag{N6-14}$$

given that $\Delta_2(t)$ remains finite in this limit. By substituting equation (N6-6) into the Laplace transform of equation (N6-14), we obtain

$$\begin{aligned}\hat{\Delta}_4(s) &= \left(1 + \frac{2}{d}\right)2s\left[\varepsilon^2 \hat{\mu}_1(s)\right]^2 \left[1 + s\eta_{p_2}^2 \hat{\phi}_{p_2}(s)\right], \\ &= \left(1 + \frac{2}{d}\right)2s\hat{\Delta}_2(s)^2 \left[1 + s\eta_{p_2}^2 \hat{\phi}_{p_2}(s)\right],\end{aligned} \tag{N6-15}$$

where the second equality follows from equation (N5-9). Equation (N6-15) is equivalent to equation (2b), because $\eta_{p_2}^2 \phi_{p_2}(t)$ in equation (N6-15) is the same as $C_{\mathcal{D}}(t)$ in equation (N6-11) for this model with the rate kernel, $\hat{\kappa}_\Gamma(s)$, given by equation (N6-4). This can be easily confirmed by substituting equation (N6-4) into equation (28). In the long-time limit as well, the NGP calculated with equation (N5-10) is the same as the NGP calculated with equation (N6-14), that is, $\alpha_2(t) \cong \Theta_N(t)[= Q_N(t)/\langle N(t) \rangle]$, which is demonstrated in Supplementary Fig. 2f.

Finally, we provide a detailed simulation algorithm used to generate the simulation results in Supplementary Fig. 2[8]. Every stochastic trajectory begins with the enzyme-substrate association step $(E + S \to ES)$. The units of length and time are chosen as $\varepsilon$ and $\langle t_{ES} \rangle$, respectively. Time elapsed for each enzyme-substrate association event $(E + S \to ES)$ is sampled from $\varphi_1^0(\tilde{t}) = xe^{-\tilde{t}x}$. Lifetime of an ES complex is then sampled from $\varphi_{ES}(\tilde{t}) = e^{-\tilde{t}}$. The fate of an ES complex between the dissociation reaction ($E + S \leftarrow ES$) and the catalytic reaction ($ES \to E + P$) is probabilistically chosen using the enzyme state $\Gamma_i$-dependent probability, $p_{2,i}(= p_2(\Gamma_i))$, of a catalytic reaction. A uniform random number is generated between 0 and 1, and if it is smaller than $p_{2,i}(t)[= p_2(\Gamma_i(t))]$ at that time, a catalytic reaction occurs, and the model motor protein jumps to one of two adjacent sites, showing no bias between either site. Otherwise, the ES complex is dissociated into a free enzyme and a substrate, and the motor protein does not move. Either case is followed by another round of enzyme-substrate association reactions. A stochastic realization of $p_{2,i}(t)$ is generated, irrespective of the enzymatic reaction process. An initial value, $p_2(0)$, is sampled with the stationary weight, 1/3, for each state. Lifetimes of $\Gamma_1$ and $\Gamma_3$ enzyme states are sampled from $k_{12}e^{-k_{12}t}$ and $k_{23}e^{-k_{23}t}$, respectively. As shown in Fig. 2a, an enzyme at both states changes to enzyme state $\Gamma_2$. Then the lifetime of enzyme state $\Gamma_2$ is stochastically determined by sampling $t_{12}$ and $t_{23}$ from $k_{12}e^{-k_{12}t}$ and $k_{23}e^{-k_{23}t}$ and by choosing whichever is smaller, or $\min(t_{12}, t_{23})$. If

$\min(t_{12}, t_{23}) = t_{12}$, enzyme state $\Gamma_2$ changes to state $\Gamma_1$ at the end of its lifetime; otherwise, it changes to state $\Gamma_3$. The values of $(p_{2,1}, p_{2,2}, p_{2,3})$, $x$, and $k_{12}\langle t_{ES}\rangle$ are given by $(1, 0.6, 0.3)$, 1, 0.1, respectively.

**SUPPLEMENTARY FIGURES**

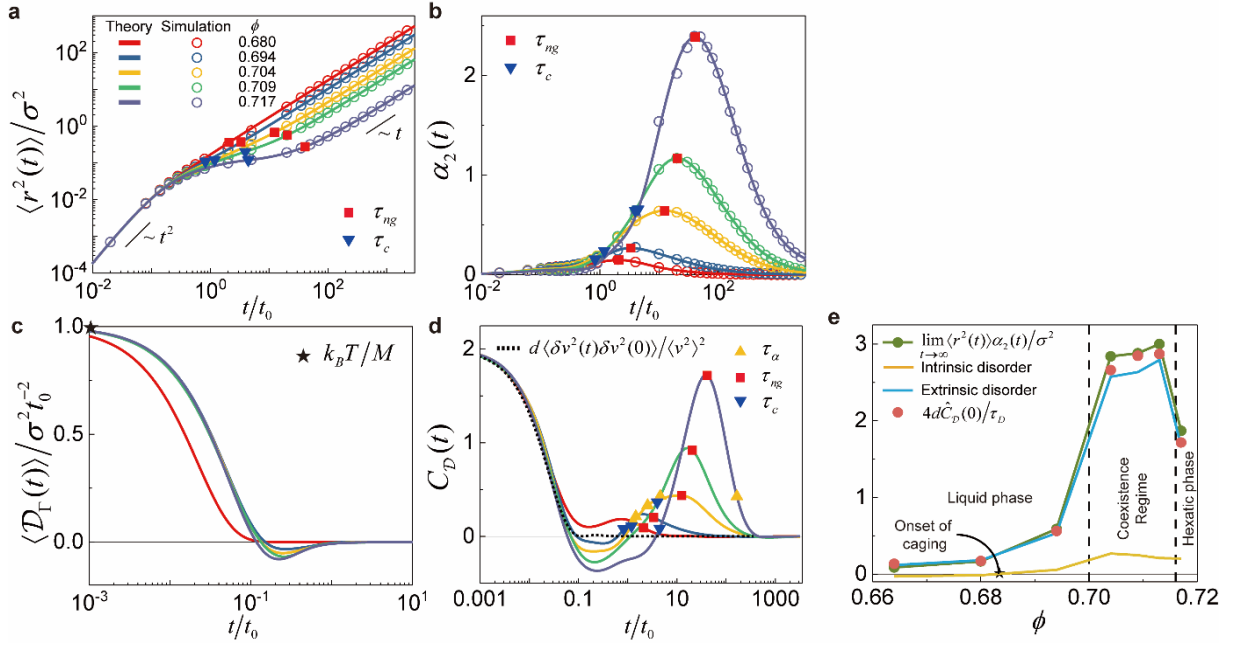

**Supplementary Figure 1. Quantitative analysis of the mean square displacement and the non-Gaussian parameter for dense hard-disc fluids. a, b,** MSDs and NGPs as functions of time at various area fractions, $\phi$. The solid lines respectively represent the best least-square fits of equation (5) and a linear combination of three or four Gaussian-shaped functions to the simulation results (open circles). **c,** Averaged diffusion kernel, or equivalently the velocity autocorrelation function obtained from the second-order time derivatives of the best MSD fits given in **a**. **d,** Diffusion kernel correlation function, $C_\mathcal{D}(t)$, calculated using equations (2a) and (2b) (see Methods). The dotted line represents the mean-scaled correlation function of squared speed at $\phi = 0.717$. In **a**, **b**, and **d**, the navy-blue triangles and the red squares represent the caging times, $\tau_c$, and the NGP peak times, $\tau_{ng}$, respectively. **e,** Total disorder, $\lim_{t\to\infty}\langle r^2(t)\rangle\alpha_2(t)$, scaled by a hard disk diameter squared, $\sigma^2$ (green circles) and its two components: intrinsic disorder and disorder (yellow and cyan lines). Extrinsic disorder estimated from the difference between the total disorder and intrinsic disorder is in good agreement the value of $4d\hat{C}_\mathcal{D}(0)/\tau_D$ (red circles) directly calculated from $C_\mathcal{D}(t)$ in **d**.

Extrinsic disorder varies both largely and non-monotonically upon phase changes while intrinsic noise does not.

\* For the two lowest densities, $\phi = 0.664$ and $0.680$ in the dense hard-disc system, we assume no bound modes in equation (5) because the minimum values of the non-Fickian coefficient, $\nu(t)\left(= d\ln\Delta_2(t)/d\ln t\right)$, are estimated to be unity, indicating an absence of any noticeable caging feature in the MSD.

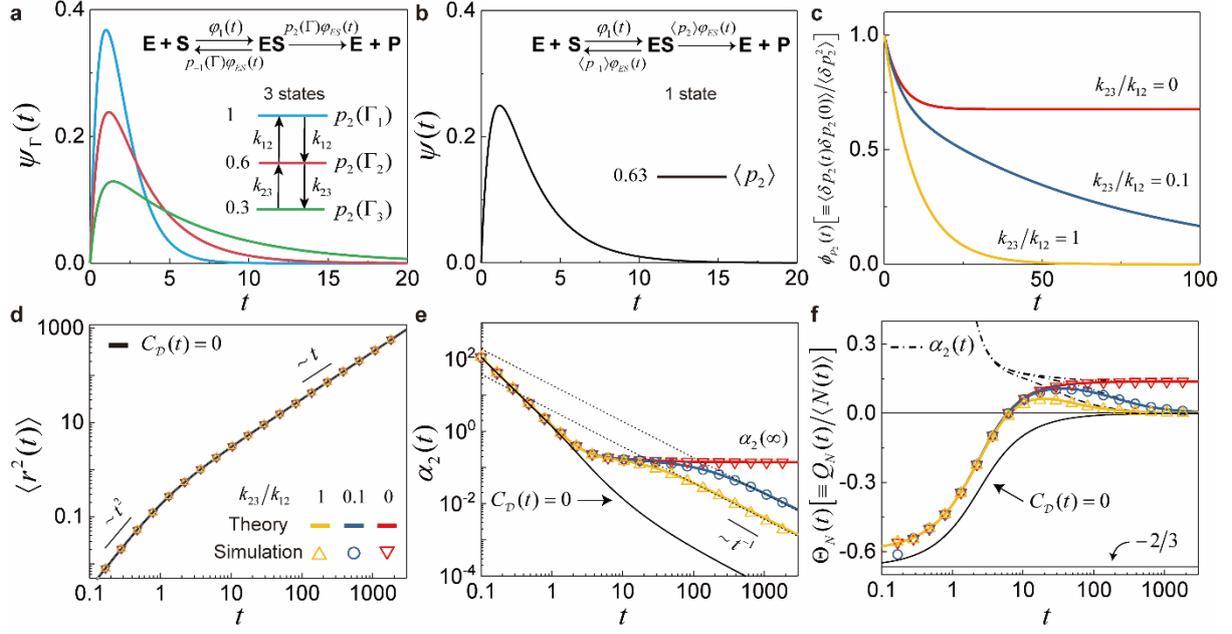

**Supplementary Figure 2. Random walk of a motor protein with state-dependent catalytic activity. a,** State-dependent waiting time distribution, $\psi_\Gamma(t)$, between successive enzymatic reactions for the three-state motor protein model. In the event of an enzyme reaction, the motor protein performs an unbiased random jump to one of the two adjacent positions. $\varphi_{1(ES)}(t)$, $S$, and $P$ represent an enzyme, substrate, and product, respectively. $\varphi_1(t)$ denotes the distribution of time required to complete an enzyme-substrate association reaction, given that the association reaction begins at time 0. $\varphi_{ES}(t)$ denotes the lifetime distribution of the enzyme-substrate complex $p_{2(-1)}(\Gamma)$. $p_2(\Gamma)$ or $k_{ij}$ denotes the probability of catalytic reaction or dissociation of the enzyme-substrate complex at state $\Gamma$. $p_2(\Gamma)$ has different values depending on the state. $k_{ij}$ or $k_{ji}$ denotes the transition rate between $\Gamma_i$ and $\Gamma_j$. All states are equally probable. **b,** Averaged waiting time distribution $\psi(t)$. $\langle p_2 \rangle$ denotes the mean value of $p_2(\Gamma)$. **c,** normalized time correlation functions, $\phi_{p_2}(t)\left[=\langle \delta p_2(t)\delta p_2(0)\rangle/\langle \delta p_2^2\rangle\right]$, of $p_2(\Gamma)$, **d,** mean square displacement (MSD), **e,** non-

Gaussian parameter (NGP), **f,** non-Poisson component $\Theta_N(t)[=Q_N(t)/\langle N(t)\rangle]$, of the relative variance in the jump event number, $N$, for three different values of $k_{23}/k_{12}$. In **d**, **e**, and **f**, the coloured lines represent the theoretical results obtained by using $\psi_\Gamma(t)$ given in **a**, which are in excellent agreement with the simulation results (open symbols) (see Supplementary Note 6). In **d**, **e**, and **f**, the black lines represent the results in the limiting case where $C_D(t)=0$, which are obtained by using $\psi(t)$ given in **b** for each quantity. In **f**, the dash-dot lines represent the NGPs given in **e**. The NGP and $\Theta_N(t)$ are equal at long times. For the non-ergodic case with $k_{23}/k_{12}=0$ (red lines and symbols), the NGP and $\Theta_N(t)$ do not vanish in the long-time limit but have the same finite value as $\alpha_2(\infty)=\eta_{p_2}^2\phi_{p_2}(\infty)$. For the ergodic case with $k_{23}/k_{12}>0$, both vanish following inverse time dependence.

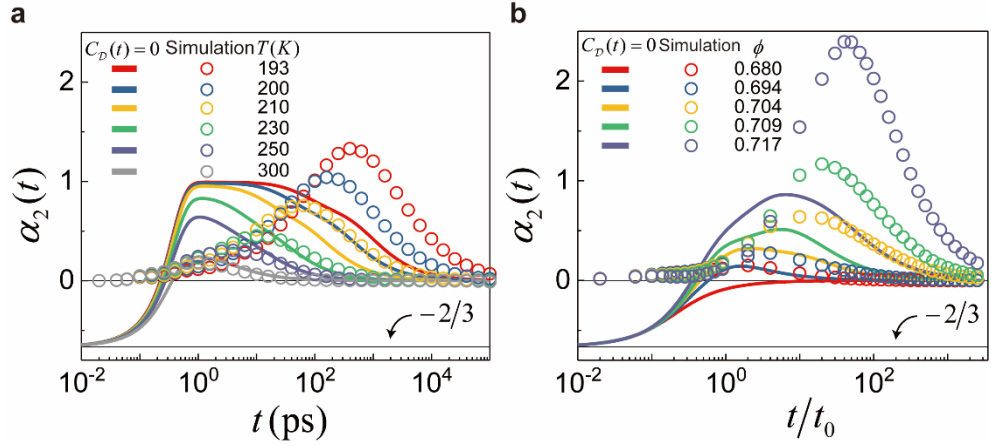

**Supplementary Figure 3. NGP comparsion between the environment-invariant model and simulation results.** The solid lines represent the NGP in the limiting case where $C_\mathcal{D}(t)=0$ for **a,** the TIP4P/2005 water system and **b,** the dense hard-disc system. In this case, the NGP is calculated by using the best fit of equation (5) to the MSD data and equation (2b) with $C_\mathcal{D}(t)=0$. Circles represent the simulation results for the NGP. This figure shows that the environment-invariant model with $C_\mathcal{D}(t)=0$ cannot quantitatively explain the time-dependence of NGP for either system; the value of the NGP for the environment-invariant model approaches $-2/3$ in the short-time limit, whereas the value of NGP from the simulation approaches zero (see Supplementary Note 4).

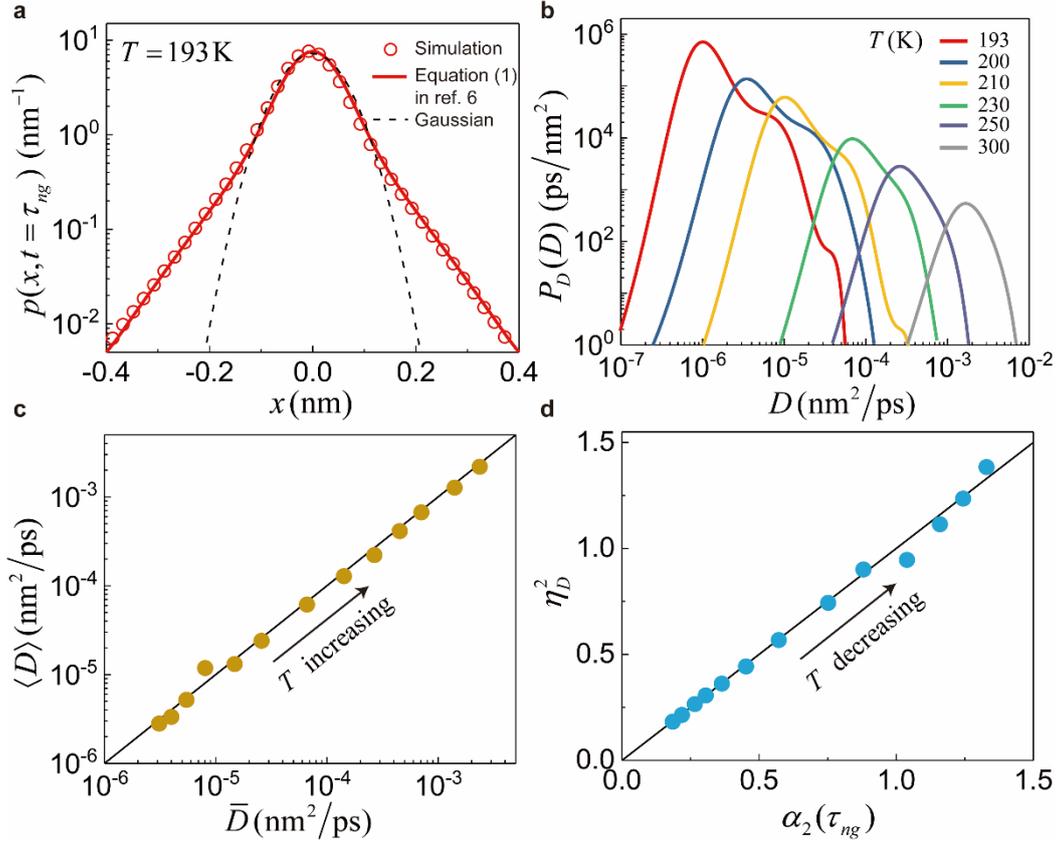

**Supplementary Figure 4. Heterogeneous diffusion coefficients extracted from the diplacement distribution for the TIP4P/2005 water system. a,** The one-dimensional displacement distribution, $p(x,t)$, at the NGP peak time, $t = \tau_{ng}$, for the TIP4P/2005 water system at temperature, $T = 193$ K. The displacement distribution (circles) obtained from the MD simulation is highly non-Gaussian with exponential-like tails, which largely deviates from a Gaussian distribution (dashed line) fitted to the central portion. **b,** The distribution, $P_D(D)$, of diffusion coefficients ($D$) for the TIP4P/2005 water system at various temperatures. The distribution of diffusion coefficients is obtained by inverting the following equation, $p(x,t = \tau_{ng}) = \int_0^\infty dD P_D(D) e^{-x^2/4D(\tau_{ng}+t_s)} / \left[4\pi D(\tau_{ng}+t_s)\right]^{1/2}$ [9]. Here, $p(x,t = \tau_{ng})$ indicates the displacement distribution from the MD simulation, which is interpreted as the average of a single Gaussian displacement distribution at $t = \tau_{ng} + t_s$ over the stationary distribution of diffusion coefficients. This interpretation corresponds to the short-time limit of the

displacement distribution in the SD model[2, 7, 8]. The value of the shifting time, $t_s$, is determined by identifying the long-time MSD, given in equation (3a), as $2d\langle D\rangle(t+t_s)$ (see Methods). The solid line in **a** represents the displacement distribution reconstructed with the distribution of diffusion coefficients at $T = 193$ K, which is well superimposed on the simulation result. **c,** The mean diffusion coefficient, $\langle D \rangle$, calculated using the distribution of diffusion coefficients given in **b** is nearly the same as the mean diffusion coefficient, $\bar{D}$, obtained from the Fickian regime of the MSD given in Fig. 2a. **d**, The relative variance, $\eta_D^2$, of the diffusion coefficient, calculated by using the distribution of diffusion coefficients given in **b**, is also essentially the same as the peak height, $\alpha_2(\tau_{ng})$, of the NGP.

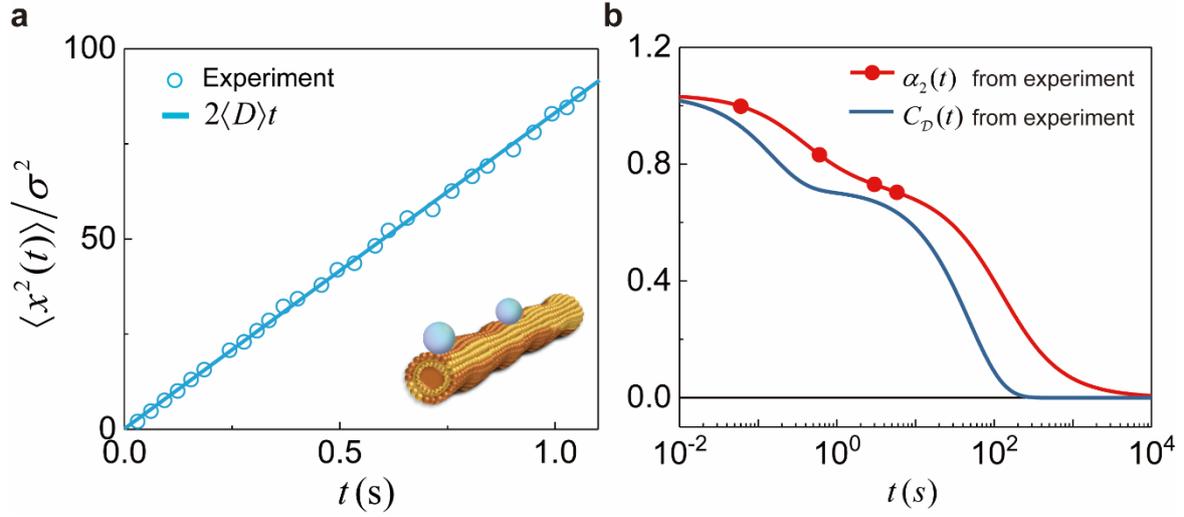

**Supplementary Figure 5. MSD, NGP and diffusion kernel correlation function extracted from the analysis of the experimental data for colloidal beads on lipid tubes. a,** MSD scaled by bead diameter squared, $\sigma^2$, and **b,** NGP, $\alpha_2(t)$, and diffusion kernel correlation function, $C_D(t)$, for colloidal beads moving on lipid tubes. In **a**, circles and solid lines represent the experimental data and the best fit of $\langle x^2(t)\rangle = 2\langle D\rangle t$ to the data, respectively. The value of $\langle D\rangle/\sigma^2$ is found to be 41.5 s$^{-1}$. In **b**, the red and blue solid lines respectively represent equations (45) and (52b) with the values of the adjustable parameters given in Table 2, which are determined by fitting equation (53) to the experimental data for the displacement distributions at four different times (see Methods).

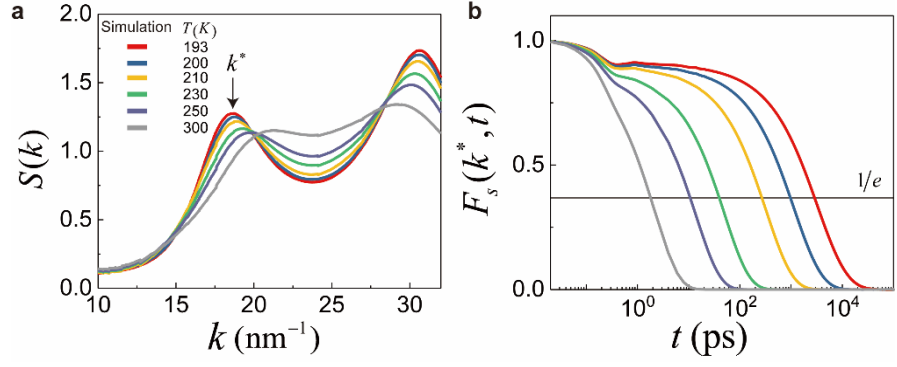

**Supplementary Figure 6. Static structure factor and self-part of intermediate scattering function for the TIP4P/2005 water system. a,** The oxygen-oxygen static structure factor, $S(k) = \frac{1}{N}\left\langle \sum_{i=1}^{N}\sum_{j=1}^{N} e^{i\mathbf{k}\cdot(\mathbf{r}_i - \mathbf{r}_j)} \right\rangle$ for the TIP4P/2005 water system. Here, $k$, $N$, and $\mathbf{r}_i$ denote the magnitude of a scattering vector, $\mathbf{k}$, the number of water molecules, and the position vector of the central oxygen atom in the $i$th water molecule, respectively. $S(k)$ is calculated by using the method given in ref. [12]. $k^*$ indicates the first peak position of $S(k)$. **b,** The self-part, $F_s(k,t)\left(=\langle e^{i\mathbf{k}\cdot[\mathbf{r}(t)-\mathbf{r}(0)]}\rangle\right)$, of the intermediate scattering function at $k = k^*$ for the TIP4P/2005 water system. $\mathbf{r}$ denotes the position vector of the central oxygen atom in a water molecule. The alpha relaxation time, $\tau_\alpha$, is defined by $F_s(k^*, \tau_\alpha) = e^{-1}$.

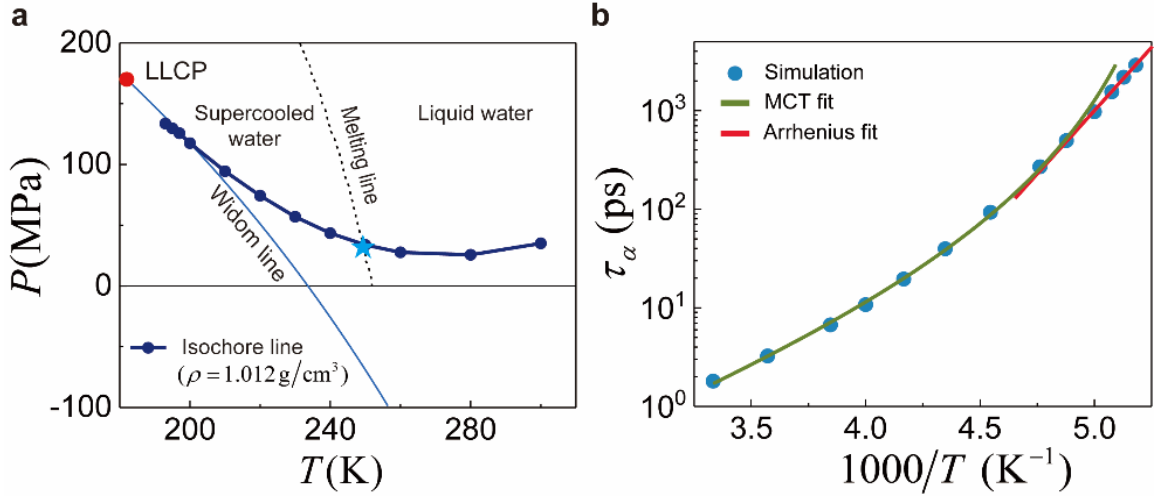

**Supplementary Figure 7. Widom line temperature and fragile-to-strong crossover. a,** The pressure-temperature phase diagram of TIP4P/2005 water. The isochore line with density, $1.012\text{g}\cdot\text{cm}^{-3}$, joining the phase points at which our MD simulation runs meets the melting line at 250 K [13] and the Widom line at 200 K. The critical pressure and temperature at the liquid-liquid critical point (LLCP) are tabulated in Table I of ref. [14]. The Widom line here is given by the pressure-temperature curve satisfying the condition that equation (3), given in ref. [14], is equal to zero. **b,** The Arrhenius plot of the alpha relaxation time, $\tau_\alpha$ (see Fig. S5). At temperatures higher than the Widom line temperature, $T_W = 200\text{K}$, the temperature dependence of $\tau_\alpha$ is well described by the mode coupling theory (MCT) prediction, i.e. $\tau_\alpha \sim (T - T_c)^{-\gamma}$ with $T_c$ denoting the MCT glass transition temperature. At temperatures lower than $T = 210\text{K}$, $\tau_\alpha$ follows the Arrhenius behavior, i.e. $\tau_\alpha \sim e^{E_a/k_B T}$ with $E_a$ denoting the activation energy for the escape of a tracer particle from a cage. The values of $T_c$, $\gamma$, and $E_a$ are respectively obtained as $T_c = 183\text{K}$, $\gamma = 3.41$, and $E_a = 49.1\text{kJ}\cdot\text{mol}^{-1}$, which are consistent with the results given in ref. [15]. The fragile-to-strong crossover between the two regimes occurs around the Widom line temperature, $T_W = 200\text{K}$.

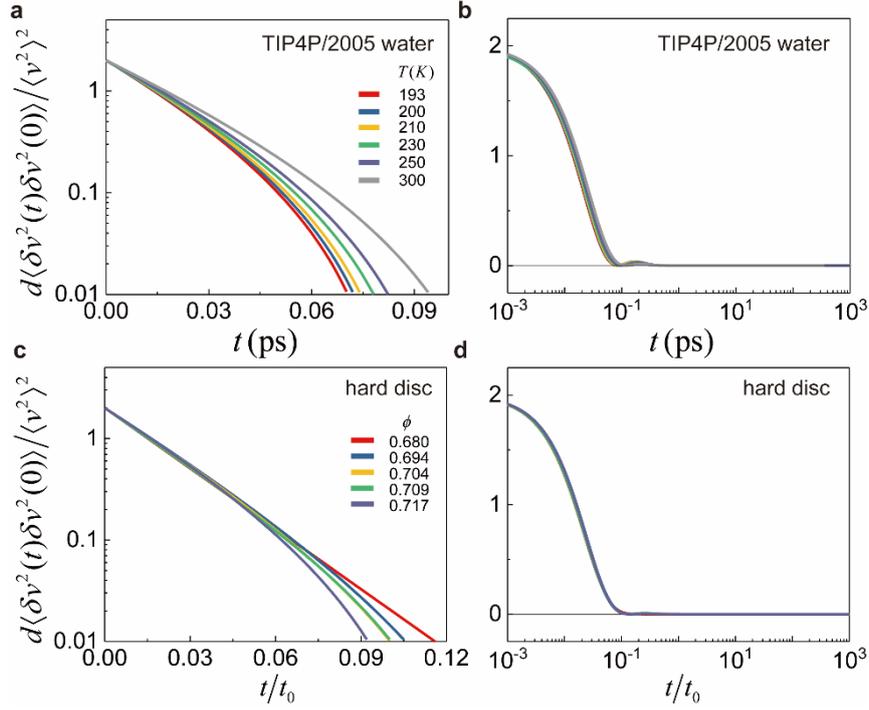

**Supplemenatry Figure 8. Time correlation function of squared speed for the TIP4P/2005 water and dense hard-disc systems.** The mean-scaled time correlation function, equation (33), of squred speed, $v^2 (= |\mathbf{v}|^2)$, **a,** at short times and **b,** over the whole time range for the TIP4P/2005 water system. **c, d**, The mean-scaled time correlation function, equation (33), of squred speed, $v^2 (= |\mathbf{v}|^2)$, at short times and over the whole time range for the dense hard-disc system. Equation (33), $d\langle \delta v^2(t) \delta v^2(0) \rangle / \langle v^2 \rangle^2$, is calculated by using equations (32) and (N4-6). Here, $\Delta_2(t)$ is given by the best fit of equation (5) to the MSD data. As shown in **a** and **c**, the time correlation function of squared speed decays faster as temperature ($T$) decreases or the area fraction ($\phi$) increases. However, as shown in **b** and **d**, the changes in the time correlation function of squared speed with temperature or density are not easily discernible in a linear scale.

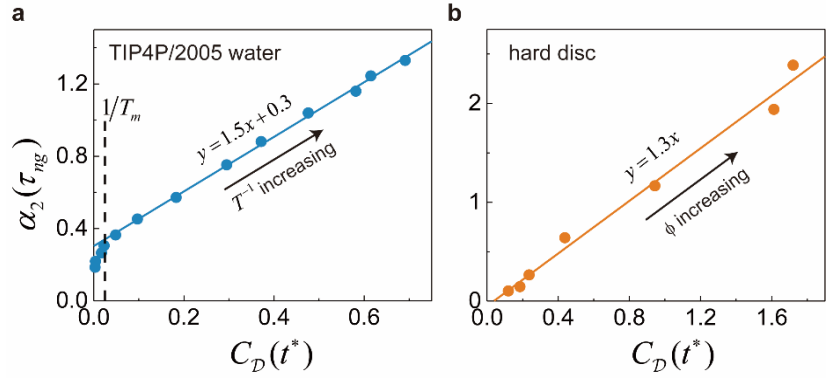

**Supplementary Figure 9. Linear relationship between peak heights of the NGP and the diffusion kernel correlation function.** The NGP, $\alpha_2(t)$, has a single peak at the peak time, $\tau_{ng}$, as shown in Figs. 2b and Supplementary Fig. 1b. The diffusion kernel correlation function, $C_\mathcal{D}(t)$, reaches its second peak at the peak time, $t^*$, as shown in Figs. 2d and Supplementary Fig. 1d. The two peak heights, $\alpha_2(\tau_{ng})$ and $C_\mathcal{D}(t^*)$, of the NGP and the diffusion kernel correlation function are found to be positively and linearly correlated with each other for **a,** the TIP4P/2005 water system and **b,** the dense hard-disc system. As inverse temperature ($1/T$) or the area fraction ($\phi$) increases, values of the peak heights also increase. The solid lines indicate simple linear fits to the data. In the case of the TIP4P/2005 water system at temperatures higher than the melting temperature, $T_m = 250\text{K}$, $\alpha_2(\tau_{ng})$ is linearly dependent on $C_\mathcal{D}(t^*)$ but with a different slope.

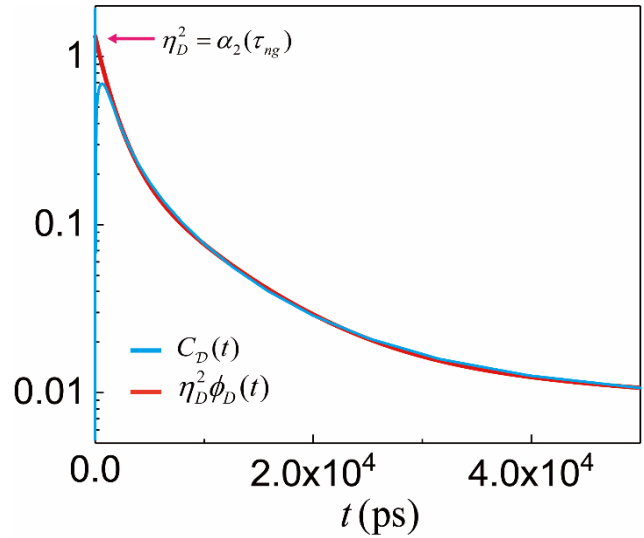

**Supplementary Figure 10. The long-time behavior of the diffusion kernel correlation function for the TIP4P/2005 water system at 193 K.** The blue solid line represents the diffusion kernel correlation funciton, $C_D(t)$, for the TIP4P/2005 water system at 193 K, which is given in Fig. 2d. The red solid line represents the best fit of equation (49c) with three modes to the time profile of $C_D(t)$ over the time range beyond the NGP peak time. As shown in this figure, the long-time profile of $C_D(t)$ shows a highly non-exponential relaxation behavior.

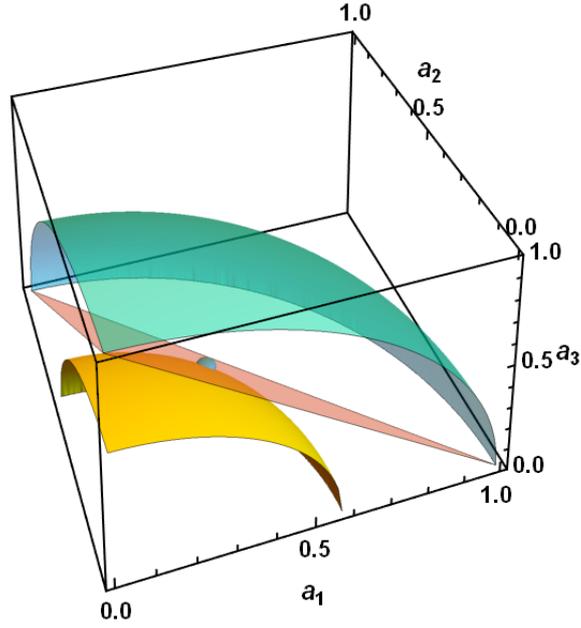

**Supplementary Figure 11. The range of $\eta_D^2$ over which the diffusion coefficient fluctuation model, equation (51) can be used.** In the model of the diffusion coefficient fluctuation, which is described by equations (48) and (51), the Ornstein-Uhlenbeck mode coefficients, $\{a_{1\leq j \leq n_a}\}$, in equation (51) should satisfy both conditions, $\sum_j a_j = 1$ and $\sum_j a_j^2 = \eta_D^2/2$, simulataneously. In other words, two surfaces, $f(\mathbf{a}) = \sum_j a_j - 1 = 0$ and $g(\mathbf{a}) = \sum_j a_j^2 - \eta_D^2/2 = 0$ in the $n_a$-dimensional coordinate space spanned by positive $a_i$'s should intersect each other. Because of this condition, values of $\eta_D^2$ available for this model are limited to the following range: $2/n_a \leq \eta_D^2 \leq 2$. The gradient vector, $\nabla f(\mathbf{a}) \left[ = \sum_j \mathbf{e}_j (\partial f / \partial a_j) = \sum_j \mathbf{e}_j \right]$, where $\mathbf{e}_j$ denotes the $j$th basis vector along $a_j$ axis, intersects with the plane, $f(\mathbf{a}) = 0$, at the point, $(n_a^{-1}, n_a^{-1}, \cdots, n_a^{-1})$. Because $\nabla f(\mathbf{a})$ is perpendicular to $f(\mathbf{a}) = 0$, the minimum distance between the coordinate origin and the plane is given by the distance between the coordinate origin and $(n_a^{-1}, n_a^{-1}, \cdots, n_a^{-1})$, that is,

$(\sum_j n_a^{-2})^{1/2} = n_a^{-1/2}$, which is, simultaneously, the minimum radius the sphere, $g(\mathbf{a}) = 0$, can take to intersect with the plane; accordingly, $n_a^{-1/2} \leq \eta_D/2^{1/2}$ or $2/n_a \leq \eta_D^2$. On the other hand, the maximum radius the sphere, $g(\mathbf{a}) = 0$, can take to intersect with the plane is unity; accordingly, $\eta_D/2^{1/2} \leq 1$ or $\eta_D^2 \leq 2$. There are, in total, $n_a$ intersecting points between the plane and $n_a$ coordinate axes, explicitly, $(1,0,\cdots,0)$, $(0,1,\cdots,0)$, …, $(0,0,\cdots,1)$. These points are also intersecting points between the plane and the unit sphere, $\sum_j a_j^2 = 1$. The figure shows an explicit example when $n_a = 3$. The yellow, pink, and pale-green surfaces respectively represent $a_1^2 + a_2^2 + a_3^2 = \eta_D^2/2 = 1/n_a = 1/3$, $a_1 + a_2 + a_3 = 1$, and $a_1^2 + a_2^2 + a_3^2 = 1$. The cyan sphere indicates the intersecting point between the yellow and pink surfaces, explicitly, $(\tfrac{1}{3}, \tfrac{1}{3}, \tfrac{1}{3})$.

# SUPPLEMENTARY REFERENCES